\documentclass[aps,prd,amsmath,showpacs]{revtex4}
\usepackage{epsf}
\newcommand{\fr}[2]{\frac{#1}{#2}}
\newcommand{\Ref}[1]{(\ref{#1})}
\newcommand{\be}{\begin{equation}}
\newcommand{\ee}{\end{equation}}
\newcommand{\bn}{\begin{eqnarray}}
\newcommand{\en}{\end{eqnarray}}
\newcommand{\bd}{\begin{displaymath}}
\newcommand{\ed}{\end{displaymath}}
\newcommand{\bnn}{\begin{eqnarray*}}
\newcommand{\enn}{\end{eqnarray*}}
\newcommand{\adb}{\allowdisplaybreaks }
\newcommand{\bs}{\begin{subequations}}
\newcommand{\es}{\end{subequations}}
\newcommand{\ak}{a_k}

\newcommand{\rk}{r_{\hspace{-0.2em}k}}
\begin{document}

\title{Semiclassical wormholes}

\author{Nail R. Khusnutdinov}
\email{e-mail: nk@dtp.ksu.ras.ru} \affiliation{Department of
Physics, Kazan State Pedagogical University, Mezhlauk 1, Kazan ,
420021, Russia}\date{\today}

\begin{abstract}
Smooth-throat wormholes are treated on as possessing quantum
fluctuation energy with scalar massive field as its source. Heat
kernel coefficients of the Laplace operator are calculated  in
background of the arbitrary-profile throat wormhole with the help
of the zeta-function approach. Two specific profile are
considered. Some arguments are given that the wormholes may exist.
It serves as a solution of semiclassical Einstein equations in the
range of specific values of length and certain radius of
wormhole's throat and constant of non-minimal connection.
\end{abstract}

\maketitle
\section{Introduction}\label{Intro}

Great interest towards the space-time of wormholes dates back
at least to 1916 \cite{Fla16}. Subsequent activity was initiated by
both classical works of Einstein and Rosen in 1935 \cite{EinRos35} in the
context of black hole space-time structure and the later series of
works by Wheeler in 1955 \cite{Whe55} with his excellent idea to
create everything from nothing. The more recent interest in the
topic of wormholes has been rekindled by the works of Morris and
Thorne \cite{MorTho88} and Morris, Thorne, and Yurtsever
\cite{MorThoYur88} who made use of the concept of wormhole in scientific
discussion of "time machine". These authors constructed and
investigated a class of objects they referred to as "traversable
wormholes". Their work led to a flurry of activity in wormhole
physics \cite{VisserBook}.

It is well-known that the central problem of the traversable
wormholes connects with unavoidable violations of the null energy
condition. It means that the matter which should be a source of
this object has to possess some exotic properties. For this
reason the traversable wormhole can not be represented as a
self-consistent solution of Einstein equations with usual
classical matter as a source because usual matter is sure to
satisfy all energy conditions. One way out is to use quantum fields
in frameworks of semi-classical quantum gravity. The point is that
the vacuum average value of energy-momentum tensor of quantum
fluctuations may violate energy conditions. A self-consistent
wormholes in framework the semiclassical quantum gravity have been
studied in Ref. \cite{Sus92}. In our recent paper \cite{KhuSus02}
we have considered a possibility for self-consistent solution of
semi-classical Einstein equations for specific kind of wormhole --
short-throat flat-space wormhole. The model represents two
identical copies of Minkowski space with spherical regions excised
from each copy, and with boundaries of these regions to be
identified. The space-time of this model is flat everywhere except
a two-dimensional singular spherical surface. The vacuum average
of energy of quantum fluctuations of massive scalar field with
non-minimal connection serves as a source for this space-time. Due
to the fact that this space-time is flat everywhere a complete set
of wave modes of the massive scalar field can be constructed and
ground state energy can be calculated. In the paper we present a
calculation of full energy of quantum fluctuations rather then
energy density and use the Einstein's equations with quantum source only,
without classical contribution. We found that the energy of
fluctuations as a function of radius of throat $a$ may possess a
minimum  if the non-minimal connection constant $\xi > 0.123$.
Utilization of the Einstein equations at the minimum gives the stable configurations
of the wormhole. For instance, in the case of conformal
connection, $\xi = 1/6$, we found relation between the radius $a$
of wormhole and mass $m$ of the scalar field: $am \approx 0.16$.
The Einstein equations say that the wormhole has a radius of throat
$a \approx 0.0141 l_{Pl}$ and the mass of scalar field $m \approx
11.35m_{Pl}$. Therefore, this kind of wormhole, if it exists, may
possess sub-Planckian radius of throat and it may be created by a
massive scalar field with supper-Planckian mass. Obviously, the
validity of the results obtained are restricted by the model taken  --
short-throat flat-space wormhole.

The goal of this paper is to consider the wormholes with more real
geometry of throat and energy of quantum fluctuations of massive
scalar field as a source of this background. The main problem in
this case has rather mathematical character. Even for simple
profile of throat it becomes impossible to obtain a full set of
solutions of radial equation in order to find the energy density
of quantum fluctuations in close form. Nevertheless, it is
possible to make some predictions about the existence of the
wormholes by considering the heat kernel coefficients
\cite{KhuSus02}. In fact, the crucial point is the existence of
the negative minimum of the zero point energy. The sufficient
condition for the zero point energy to have negative minimum is
that the heat kernel coefficients $B_2$ and $B_3$ be positive
\cite{KhuSus02}. This gives a condition for parameters of model.
More precisely, if a background is described by a parameter $\tau$
with dimension of length and the domain where the space-time is
"mainly"\/ curved is defined by this parameter, then for small
size of curved domain, $\tau\to 0$, the zero point energy shows
the following behavior
\bd
E^{ren} \approx -\fr{B_2 \ln(\tau m)^2}{32\pi^2},
\ed
and in opposite limit $\tau\to \infty$ we have
\bd
E^{ren} \approx -\fr{B_3}{32\pi^2 m^2}.
\ed

If both these conditions are satisfied one can expect that the
system will stay in minimum of energy which is characterized by
specific values of parameters of wormhole and constant of
non-minimal connection $\xi$. Next step is the
utilization of the Einstein's equations with energy-momentum
tensor of quantum fluctuations as a source. Integration over volume
the $t-t$ component of this equation gives an additional relation
between parameters of wormhole and zero point energy by using which
we obtain the size of wormhole and mass of scalar field in terms of
the Planck length and Planck mass correspondingly. At the
beginning we may expect \cite{KhuSus02} that the size of wormhole
and mass of field will be in the Planck scale. For this reason we
are interested only in finding of the domain of the wormhole's parameters
and constant non-minimal connection $\xi$ for different models of
the wormhole's profile.

The manifest expression for coefficient $B_2$ exists for arbitrary
background, but this is not the case for coefficient $B_3$. For
this reason we adopt here the zeta-regularization approach (see
Sec. \ref{ZeroPoint}), in frame of which it is possible to
calculate the heat kernel coefficients and zero point energy
itself. We pursue here another goal -- to evolve the zeta-function
approach for situations where it is impossible to find
the full set of solutions of radial equation in closed form.
We find a method to calculate the heat kernel coefficients in the
background of wormhole with arbitrary profile of throat by using the WKB
approach. Moreover, we obtain expressions for arbitrary heat
kernel coefficients and we reproduce them in manifest form up to
$B_3$ for arbitrary profile of wormhole's throat.

The organization of the paper is as follows. In Sec. \ref{ATraver}
we consider the geometry of wormhole with smooth throat. In Sec.
\ref{ZeroPoint} we discuss the method of zeta-function for
calculation of zero-point energy. The WKB approach for scalar
massive field is considered in Sec.\ref{WKB}. The heat kernel
coefficients are obtained in Sec.\ref{Heat}. We calculate them in
manifest form for arbitrary profile of throat. The specific
profiles of throat are investigated in Sec.\ref{Model1} and
\ref{Model2}. In Sec.\ref{Conc} we discuss the results obtained.
Appendix A contains some technical formulas which are rather
complicated to reproduce them in the text.

We use units $\hbar = c = G = 1$. The signature of the space-time,
the sign of the Riemann and Ricci tensors, are the same as in the
book by Hawking and Ellis \cite{HawEllBook}.

\section{A traversable wormhole with smooth throat}\label{ATraver}
The metric of space-time of wormhole which is under consideration
has the form below
\be\label{Metric}
ds^2 = -dt^2 + d\rho^2 + r^2(\rho)( d\theta^2 + \sin^2\theta
d\varphi^2).
\ee
The radial variable $\rho$ changes from $-\infty$ to $+\infty$. In
the paper we restrict ourself by wormholes with symmetric throat
which means that $r(-\rho) = r(+\rho)$. The radius $a$ of throat
is defined as follows $a=r(0)$. We suppose that far from the
wormhole's throat the space-time becomes Minkowskian, that is
\bd
\lim_{\rho\to\pm\infty} \fr{r^2(\rho)}{\rho^2} = 1.
\ed
The non-zero components of the Ricci tensor and scalar curvature
have the following form
\bnn
\mathcal{R}^\rho_\rho &=& -\fr{2r''}{r}, \\
\mathcal{R}^\theta_\theta = \mathcal{R}^\varphi_\varphi &=& -\fr{-1+r'^2 + r r''}{r^2},\\
\mathcal{R} &=& -\fr{2(-1+r'^2 + 2r r'')}{r^2}.
\enn
The energy-momentum tensor corresponding to this metric has
diagonal form from which we observe that the source of this metric
possesses the following energy density and pressure:
\bnn
\varepsilon &=& -\frac{-1 + r'^2 + 2rr''}{8\pi r^2},\adb\\
p_\rho &=&\frac{-1 + r'^2}{8\pi r^2},\adb\\
p_\theta &=& p_\varphi = \frac{r''}{8\pi r}.
\enn

\begin{figure}
\begin{center}
\epsfxsize=4truecm\epsfbox{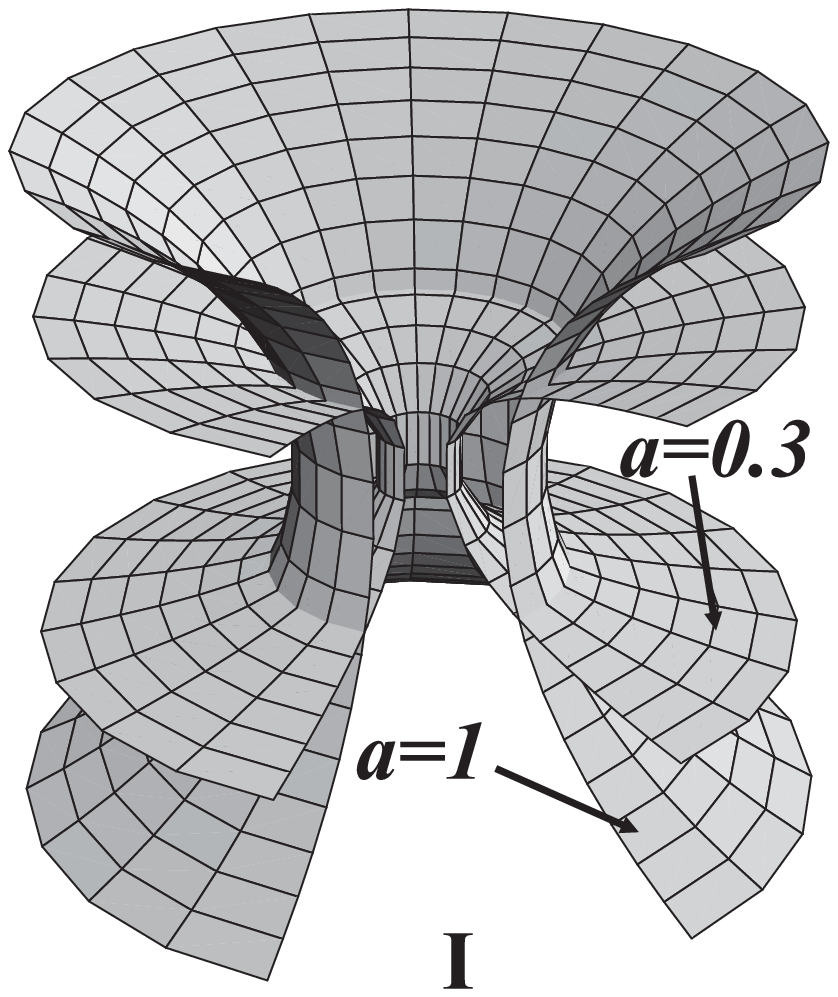}%
\epsfxsize=4truecm\epsfbox{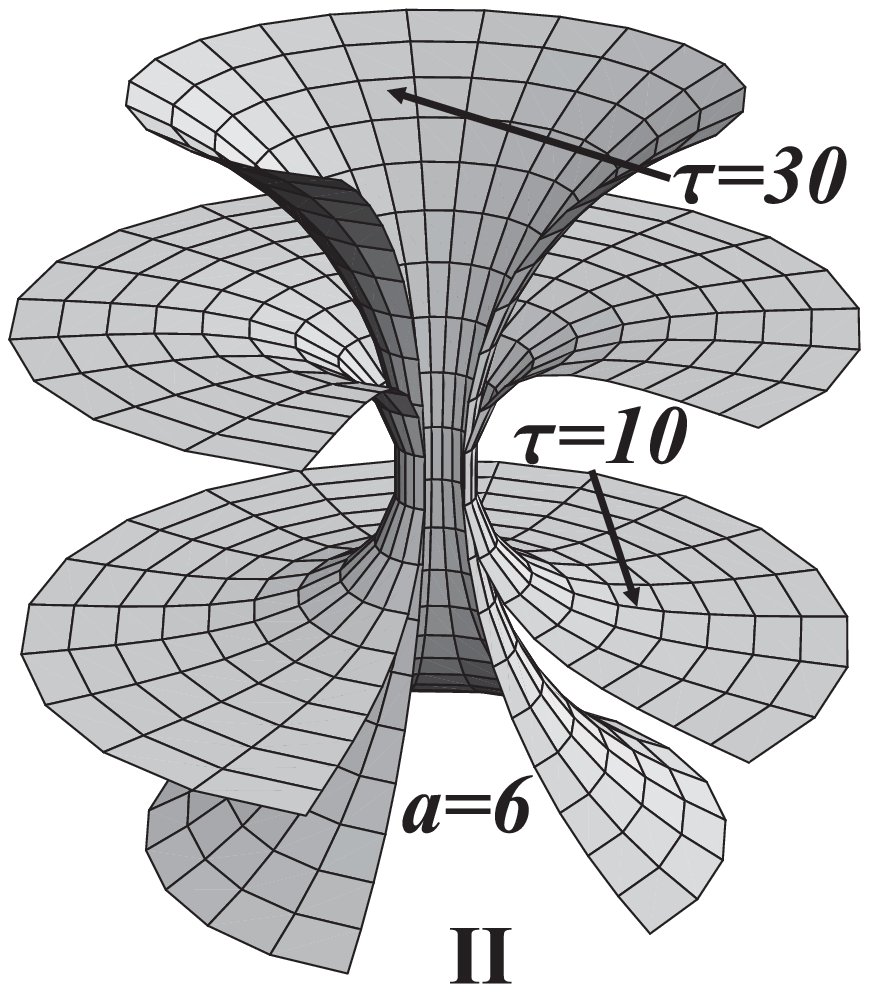}%
\epsfxsize=4truecm\epsfbox{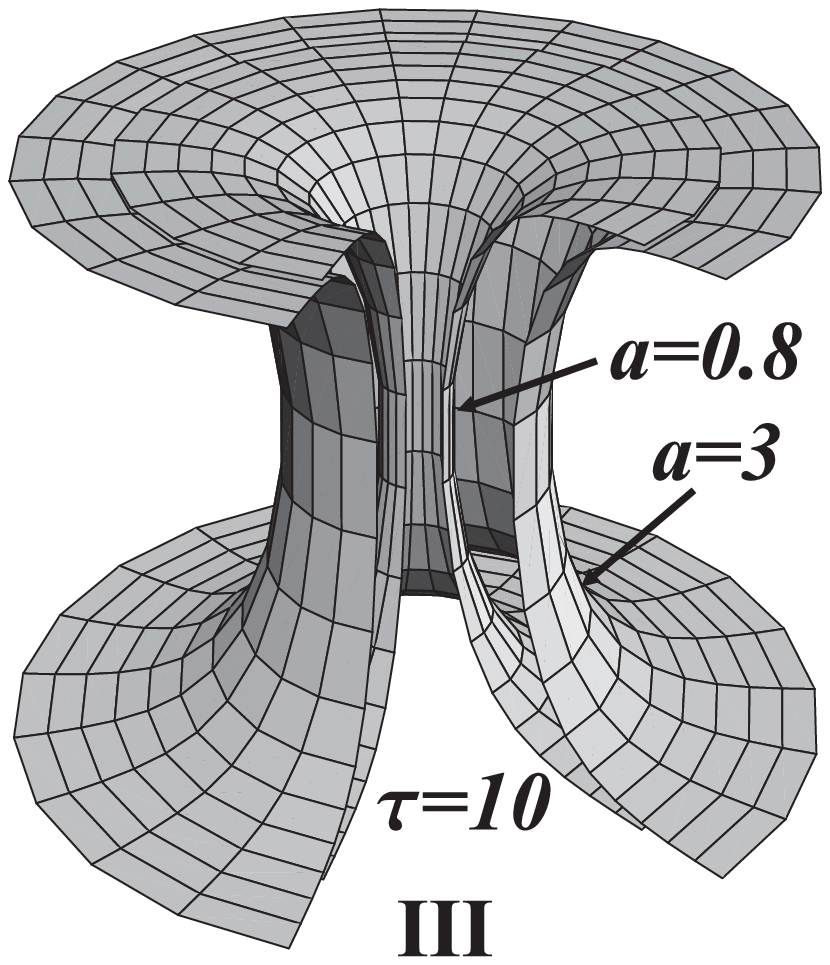}%
\epsfxsize=4truecm\epsfbox{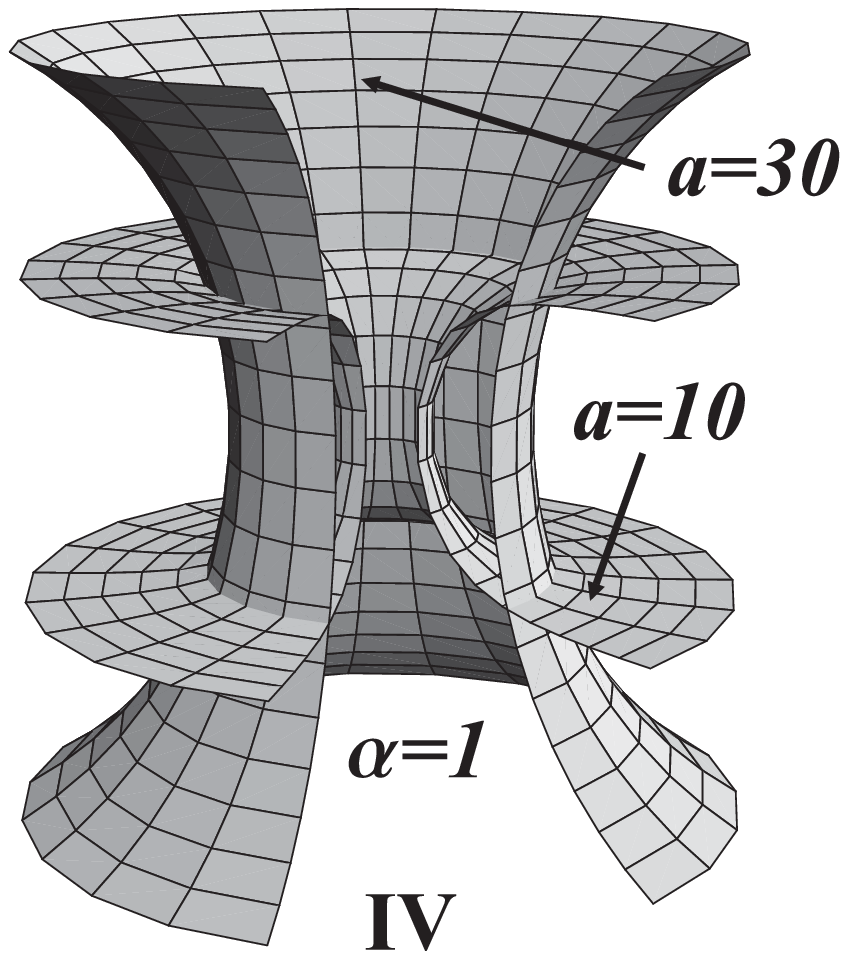}
\end{center} \caption{First figure (I) represents the section
$t=const,\theta = \pi/2$ of wormhole's space-time with profile
function $r(\rho)=\sqrt{\rho^2 + a^2}$ for two different values of
radius of the throat. Three next figures illustrate the wormhole
with profile of throat $r(\rho) = \rho \coth(\rho/\tau) - \tau +
a$. Figure (II) and (III) illustrate that the $a$ and $\tau$ are
radius and length of throat, accordingly. In last figure two
wormholes with different $a$ but with the same ratio radius and
length of throat are depicted. It is seen that the parameter $a$
characterizes the "size"\/ of wormhole and $\alpha$ describes the
"form"\/ of wormhole.} \label{worms}
\end{figure}

In the paper we obtain general formulas for space-time
\Ref{Metric} with arbitrary symmetric function $r(\rho)$ obeying
above Minkowskian condition. Two specific kinds of throat's
profile will be considered. In the first model the profile of
throat has the following form:
\be\label{First}
r(\rho) = \sqrt{\rho^2 + a^2},
\ee
where $a$ is radius of throat which characterizes wormhole's size.
The embedding into the three dimensional Euclidean space of the
section of the space-time by surface $t=const, \theta = \pi/2$ is
plotted in Fig.\ref{worms}(I) for two different values of radius
of throat. In Euclidean space with cylindrical coordinates
$(r,\varphi,z)$ this surface may be found in parametric form from
relations: $r=r(\rho),z'(\rho) = \sqrt{1-r'^2}$. In this
background there is the only nonzero component of the Ricci tensor
which reads
\bd
{\cal R}^\rho_\rho = -\fr {2 a^2}{(\rho^2 + a^2)^2}.
\ed

The second model has been considered in Ref.\cite{Sus01} and it is
characterized by the following profile of throat
\be\label{Second}
r(\rho) = \rho \coth (\fr\rho\tau) - \tau + a.
\ee
This model possesses a more reach structure. There are two parameters
$\tau$ and $a$. The latter parameter is radius of throat. In this
model we may introduce another parameter which may be called
length of the throat. The point is that the function $r(\rho)$
turns into linear function of $\rho$ starting from distance $\rho
> \tau/2$ and the space-time becomes approximately Minkowskian. Therefore,
the length of throat $l=\tau$. Using new variables $y=\rho /a,\ \alpha
= \tau/a$ one rewrites the function $r$ in the form below
\bd
r(y) =a(y\coth (\fr y\alpha ) - \alpha  + 1).
\ed
The parameter $\alpha $ is the ratio of the length and radius of
throat. This parameter will play the main role in our analysis. It
allows us to consider wormholes of different form, that is with
different ratio of the radius and length of throat.

In Fig.\ref{worms}(II-IV) the sections $t=const, \theta =\pi/2$ of
this wormhole space-time are shown for different values of $a$ and
$\tau$. Namely in Fig.\ref{worms}(II) we represent two wormholes
with the same radius of throat but with different length, and vice
versa in Fig.\ref{worms}(II) we depict two wormholes with the same
length of throat but with different radii of throat. In last
picture Fig.\ref{worms}(IV) two wormholes with the same ratio of
length and radius of throat, but with different values of throats'
radii are depicted. Therefore the size of wormhole with the same
ratio of length and radius throat is managed by parameter $a$. The
parameter $\alpha $ describes wormhole's form.

\section{Zero point energy: zeta-function approach} \label{ZeroPoint}

We exploit the zeta function regularization approach
\cite{DowCri76,ZetaBook} developed in Ref. \cite{Bor95}
and calculate the zero point energy of massive scalar field
in this background with consequent using the Einstein equations.
Let us repeat some main formulas from those papers. In framework
of this approach the zero point energy
\be \label{zeta1}
E(s)=\frac12 \mu^{2s}
\sum_{j}\sum_{(n)}(\lambda_{(n),j}^2+m^{2})^{1/2-s} =  \frac12
\mu^{2s} \zeta_{\mathcal{L}}(s-\fr 12),
\ee
of scalar massive field $\Phi$ is expressed in terms of the
zeta-function
\be \label{zeta10}
\zeta_{\mathcal{L}}(s-\fr 12) =
\sum_{j}\sum_{(n)}(\lambda_{(n),j}^2 + m^{2})^{1/2-s}
\ee
of the Laplace operator $\mathcal{L} = - \triangle + m^2 + \xi
\mathcal{R}$. Here  $\triangle = g^{kl} \nabla_k \nabla_l$ is the
three dimensional operator. The eigenvalues
$\lambda_{(n),j}+m^{2}$ of operator $\mathcal{L}$ are found from
boundary condition which looks as follows
\be \label{ConditionPsi}
\Psi_{(n)}(\lambda, R) = 0,
\ee
where $R$ denotes some boundary parameter. The solutions $\lambda
= \lambda_{(n),j}$ of this equation depend on the numbers $(n)$,
and additionally they have the index $j=1,2,\dots$, which
numerates  the solutions of the boundary equation. Therefore, the
zeta-function is a sum of expressions which depend on zeros of
function $\Psi_{(n)}$. Next, according to Ref.\cite{Bor95} we
convert the series over $j$ in zeta function to integral and
arrive to the formula
\be\label{MainFormulaEnergy}
E(s) = - \fr 12 \mu^{2s} \sum_{(n)} \fr{\cos \pi s}{\pi}
\int_m^\infty d k(k^2 - m^2)^{1/2 - s} \fr\partial {\partial k}
\ln \Psi_{(n)} (ik, R),
\ee
where the function $\Psi_{(n)}$  in imaginary axes appears.

The expression \Ref{MainFormulaEnergy} is divergent in the limit
$s\to 0$ we are interested in. For renormalization we
subtract from $E(s)$  all terms $E^{div}(s)$ which will survive in
the limit $m\to \infty$:
\bd
E^{div}(s) = \lim_{m\to \infty}E(s)
\ed
and define the renormalized energy as follows:
\be\label{ErenDef}
E^{ren} = \lim_{s\to 0} (E(s) - E^{div}(s)).
\ee
Because the pole structure of zeta-function does not depend on the
value of parameters, it is obvious that in the limit $m\to \infty$
the divergent part will have the structure of DeWitt-Schwinger
expansion, which has the following form
\bn\label{ehk}
E^{div} (s) &=& \fr 12\left(\fr\mu m\right)^{2s} \fr
1{(4\pi)^{3/2} \Gamma (s - \fr 12)} \adb\\
&\times& \left\{ B_0 m^4\Gamma (s-2) + B_{1/2}m^3\Gamma (s-\fr 32)
+ B_1 m^2 \Gamma (s-1) + B_{3/2} m \Gamma (s-\fr 12) + B_2 \Gamma
(s) \right\}, \nonumber
\en
where $B_\alpha$ are the heat kernel coefficients. In order to
extract the divergent part of the energy we use the following
procedure \cite{Bor95}. We subtract from and add to integrand the
uniform expansion of $\ln\Psi$ up to $m^0$. We denote this expansion as
$(\ln\Psi_{(n)})^{as}$. Therefore, according to this, we
represent the energy as the sum
\be\label{Division}
E(s) = E_{fin}(s) + E_{as}(s)
\ee
of finite (in the limit $s\to 0$) part
\bn
E_{fin}(s) &=& - \fr 12 \mu^{2s} \sum_{(n)} \fr{\cos \pi
s}{\pi} \int_m^\infty d k(k^2 - m^2)^{1/2 - s}\nonumber \adb\\
&\times&\fr\partial {\partial k} \left(\ln \Psi_{(n)} (ik, R) -
(\ln \Psi_{(n)}(ik, R))^{as}\right),
\en
and the remains, which will be obtained from the uniform expansion
part
\bd
E_{as}(s) = - \fr 12 \mu^{2s} \sum_{(n)} \fr{\cos \pi s}{\pi}
\int_m^\infty d k(k^2 - m^2)^{1/2 - s} \fr\partial {\partial k}
(\ln \Psi_{(n)}(ik, R))^{as}.
\ed
The last expression contains all terms which will survive in the
limit $m\to \infty$.

Taking into account the obtained expressions in Eq. \Ref{ErenDef}
we arrive at the formula
\bs\label{GenForm}
\be
E^{ren} = E_{fin} + E_{as}^{fin},
\ee
where
\bn
E_{fin} &=& E_{fin}(0) = - \fr 1{2\pi}
\sum_{(n)} \int_m^\infty d k\sqrt{k^2 - m^2}\nonumber \adb\\
&\times&\fr\partial {\partial k} \left(\ln \Psi_{(n)} (ik,
R) - (\ln \Psi_{(n)}(ik, R))^{as}\right),\label{Efin} \adb\\
E_{as}^{fin} &=& \lim_{s\to 0}(E_{as}(s) - E^{div}(s)).
\label{Easfin}
\en
\es
The divergent part $E^{div}$ is given by Eq. \Ref{ehk}.

The finite part $E_{fin}$ is calculated numerically. The second
part, in practice, is found in the following way. By using the
uniform expansion $(\ln\Psi_{(n)})^{as}$ we calculate in manifest
form the $E_{as}(s)$ and after that we take the limit $m\to
\infty$ in the expression obtained (the pole structure does not
change). All terms which will survive in this limit constitute the
DeWitt-Schwinger expansion  \Ref{ehk} which we have to subtract in
Eq. \Ref{Easfin}. This way of calculation is more preferable
because we may obtain the heat kernel coefficients in manifest
form. The calculations of heat kernel coefficients in framework of
this approach shows that the approach is suitable for both smooth
background and for manifolds with singular surfaces codimension
one \cite{KhuSus02} and two \cite{KhuBor99}, the general formulas
for which were obtained in Refs. \cite{GilKirVas01} and
\cite{Fur94b}.

From consideration above we may find the zero-point
energy for large and small size of wormhole \cite{KhuSus02}. Let
the parameter $a$ characterize the size of wormhole. In this case
the $E^{ren}/m$ is a dimensionless function and it depends on the
parameter $ma$ and some additional dimensionless parameters which
characterize the form of wormhole. For example, in first model
\Ref{First} there is the only parameter $a$ which is the radius of
wormhole's throat and it characterizes at the same time the size
of wormhole as a whole. Therefore in this model the $E^{ren}/m$
depends on $ma$ and there are no additional parameters. In the
second model \Ref{Second} there is an additional parameter $\alpha =
\tau/a$ except parameter $ma$. For this reason the dependence of
the zero point energy $E^{ren}/m$ on the mass is the same as
parameter $a$. Because for renormalization we subtracted all terms
of asymptotic over mass expansion up to $B_2$ the asymptotic $ma
\to \infty$ is the following
\be\label{Large}
\fr{E^{ren}}{m} \approx -\fr{B_3}{32\pi^2 m^3} = -\fr{b_3}{32\pi^2
(ma)^3}.
\ee
In opposite case, $ma\to 0$, the behavior of energy is defined by
coefficient $B_2$ (see also Eq. \Ref{LogTerms}):
\be\label{Small}
\fr{E^{ren}}{m} \approx -\fr{\ln(ma)^2}{32\pi^2 m} B_2 =
-\fr{\ln(ma)^2}{32\pi^2 (ma)} b_2.
\ee
Here $b_3$ and $b_2$ are dimensionless heat kernel coefficients
which may depend on the additional parameters. Therefore from
these expressions we obtain the following sufficient condition that
the zero point energy has a minimum: both $B_2$ and $B_3$ have to be
positive. An additional condition may be obtained from Einstein's
equations (see Sec.\ref{Model1},\ref{Model2}).

\section{Massive scalar field in wormholes background: the WKB
approach}\label{WKB}

We consider massive scalar quantum field in this backgrounds as a
source for this space-time. In the frameworks of the approach used one
has to find the spectrum of Laplace operator $\mathcal{L}$:
\bd
(- \triangle + \xi \mathcal{R})\Phi = \lambda\Phi.
\ed

Taking into account the spherical symmetry of problem we represent
equation in the following form
\bd
\Phi = Y_l^m(\theta,\varphi) \phi
\ed
where $Y_l^m(\theta,\varphi)$ are the spherical harmonics, $l =
0,1,2,\ldots$ and $m\in [-l,l]$. The radial part of the wave
function subjects for equation
\be\label{RadEqu}
\left(\partial^2_\rho + \fr {2r'}r
\partial_\rho  - \fr {l(l+1)}{r^2}  - \xi {\cal R}\right)\phi =
-\lambda^2 \phi.
\ee

To find the spectrum $\lambda$ we have to impose some appropriate
boundary condition. It does not matter which kind of boundary
condition will impose because in the end of calculation we will
tend this boundary to infinity. We use the Dirichlet boundary
condition in the spheres with radii $R$: $\rho = \pm R$. For
simplifying formulas we will work here with function $\zeta (s) =
m^{2s} \zeta_{\mathcal{L}} (s)$. With this notations the
regularized ground state energy reads
\bd
E(s) = \fr  12 \left(\fr \mu m\right)^{2s} \zeta (s-\fr  12).
\ed

Because we need the solution for the imaginary energy only (see
Eq. \Ref{MainFormulaEnergy}), we change the integrand variable
in radial equation \Ref{RadEqu} to imaginary axis: $\lambda
\to i \nu k$ and rescale for simplicity the radial variable: $\rho
k\to x$. Therefore we arrive at the following equation $(\nu = l +
1/2)$
\be
\ddot{\phi} + 2 \fr {\dot{\rk}}\rk \dot{\phi} - \nu^2 (1 + \fr
1{\rk^2}) \phi + (\fr  1{4\rk^2} - \xi \mathcal{R}_k)\phi = 0,
\label{RadEquDimless}
\ee
where the dot is the derivative with respect to $x$; $\rk^2 = r^2
k^2$, and $\mathcal{R}_k =\mathcal{R}/k^2$.

A general solution of radial equation \Ref{RadEquDimless} is the
superposition of two linearly independent solutions
\be
\overline{\Psi}(i \nu k,x) = C_1 \phi_1(x) + C_2 \phi_2(x).
\label{GeneralSolution}
\ee
The first function $\phi_1$ tends to infinity far from throat, for
$\rho\to \infty$, and the second one tends to zero. We consider
the behavior of functions only for one part of space-time namely,
with $\rho >0$. The behavior of the solutions in the second part
of space-time with negative $\rho$ is found as continuation of the
solutions from positive part of space-time. Now we impose the
Dirichlet boundary condition at spheres $\rho = \pm R$:
\bnn
\overline{\Psi}(i \nu k, +R) = C_1 \phi_1(+R) + C_2 \phi_2(+R)=0, \adb\\
\overline{\Psi}(i \nu k, -R) = C_1 \phi_1(-R) + C_2 \phi_2(-R)=0.
\enn
The solution of this system exists if and only if the following
condition is satisfied
\be
\Psi_l(i \nu k,R) = \phi_1(+R) \phi_2(-R) - \phi_1(-R) \phi_2(+R)
= 0. \label{MainCond}
\ee

The contribution from the second term in equation above  is
exponentially small comparing with first one in the limit $R\to
\infty$. In order to see this let us find the uniform expansion of
solutions $\phi_1$ and $\phi_2$. Moreover, we need this
expansion for renormalization and calculation of the heat kernel
coefficients. Let us represent a solution $\phi$ in the
exponential form
\be
\phi(x) =\fr 1{\sqrt{2a\nu}} e^{S(x)} \label{ExpFor}
\ee
where $a=r(0)$, and substitute it in the radial equation
\Ref{RadEquDimless}. One obtains a non-linear equation
\bd
\ddot{S} + \dot{S}^2 + 2 \fr {\dot{\rk}}\rk \dot{S} - \nu^2 (1 +
\fr 1{\rk^2}) + (\fr  1{4\rk^2} - \xi \mathcal{R}_k) = 0.
\ed

We represent now the solution in the WKB expansion form
\bd
S= \sum_{n=-1}^\infty \nu^{-n} S_n,
\ed
and substitute it in equation above. This gives the following
chain of equations
\bs\label{System}
\bn
\dot{S}_{-1} &=& \pm\sqrt{1 + \fr  1{\rk^2}}, \adb\\
\dot{S}_0 &=& -\fr  12 \fr  {\ddot{S}_{-1}}{\dot{S}_{-1}} - \fr {\dot{\rk}}{\rk}, \adb\\
\dot{S}_1 &=& -\fr  1{2\dot{S}_{-1}} \left[\ddot{S}_0 +
\dot{S}_0^2 +
2\fr {\dot{\rk}}\rk \dot{S}_0 + \fr  1{4\rk^2} - \xi \mathcal{R}_k\right], \adb\\
\dot{S}_{n+1} &=& -\fr  1{2\dot{S}_{-1}} \left[\ddot{S}_n +
\sum_{k=0}^n \dot{S}_k \dot{S}_{n-k} + 2\fr {\dot{\rk}}\rk
\dot{S}_n\right],\ n=1,2,\cdots .
\en
\es

There are two solutions of this chain corresponding to sign in the
first equation. The sign plus gives the growing (for positive
coordinate $\rho$) solution which we mark "$+$" and sign minus
gives solution which tends to zero at infinity which we mark by
sign "$-$". Therefore
\bd
\phi_1(+R) \phi_2(-R) =\fr 1{2a\nu} e^{S^+(+R)+S^-(-R)}.
\ed

To find an expansion for the sum $S^+(+R)+S^-(-R)$ we need the
following properties of function $S^\pm (x)$
\bnn
\dot{S}^{-}_{2n-1}(x) &=& -\dot{S}^{+}_{2n-1}(x),\adb\\
\dot{S}^{-}_{2n}(x) &=& + \dot{S}^{+}_{2n}(x),
\enn
and
\bnn
\dot{S}^\pm_{2n-1}(x) &=& +\dot{S}^\pm_{2n-1}(-x),\adb\\
\dot{S}^\pm_{2n}(x) &=& -\dot{S}^\pm_{2n}(-x),
\enn
where $n=0,1,\ldots$ . The first two equations are the consequence
of the structure of chain and the last two equations are due to
the symmetry of metric function $\rk(x) = \rk(-x)$.

Taking into account these properties of symmetry we have
\bn
S^+(+x)+S^-(-x) &=& \sum_{n=0}^\infty \nu^{1-2n} \left[ C_{2n -
1}^+ + C_{2n - 1}^{-}  + \int_{-x}^{+x} \dot{S}^+_{2n-1} dx \right]\label{S+-}\adb\\
&+& \sum_{n=0}^\infty \nu^{-2n} \left[ C_{2n}^+ +
C_{2n}^{-}+2 \int_{x_0}^{+x}\dot{S}^+_{2n} dx \right], \nonumber \adb\\
S^+(+x)+S^-(+x) &=& \sum_{n=0}^\infty \nu^{1-2n} \left[ C_{2n -
1}^+ + C_{2n - 1}^{-}\right]\label{S++}\adb\\
&+& \sum_{n=0}^\infty \nu^{-2n} \left[ C_{2n}^+ + C_{2n}^{-}+2
\int_{x_0}^{+x}\dot{S}^+_{2n} dx \right], \nonumber
\en
Here the $C_n$ are the constant of integration of system
\Ref{System}.

Therefore we may express the combination we need \Ref{S+-} in term of the
\Ref{S++}:
\bd
S^+(+x)+S^-(-x) = S^+(+x)+S^-(+x) + \sum_{n=0}^\infty
\nu^{1-2n}\int_{-x}^{+x} \dot{S}^+_{2n-1} dx
\ed
To find the combination $S^+(+x)+S^-(+x)$ we use the Wronskian
condition. Because these solutions are independent, they obey
to equation ($\ak = a k$)
\bd
W(\phi_1(x),\phi_2(x)) = \fr {k}{\rk^2}.
\ed

The origin of this relation is the following. Suppose we try to
find the scalar Green function of the Klein-Gordon equation:
\be
\left(g^{\mu \nu} \nabla_\mu\nabla_\nu - m^2 - \xi {\cal R}\right)
G(x,x') = \fr {\delta^4(x,x')}{\sqrt{-g(x)}} \label{KleGorGre}
\ee
in background \Ref{Metric}. It is very easy to extract time and
angular dependence of the Green function
\bd
G(x,x')= \int_{-\infty}^{+\infty} \fr{d\omega}{2\pi}
\sum_{l=0}^\infty \sum_{m=-l}^l Y_l^m(\theta,\varphi)
Y_l^{-m}(\theta',\varphi') e^{-i\omega (t-t')}\phi(\rho,\rho'),
\ed
and we arrive to equation for radial part of Green function which
reads $(\lambda^2 = \omega^2 - m^2)$
\bd
\left\{\partial^2_\rho + \fr {2r'}r \partial_\rho + \lambda^2 -
\fr {l(l+1)}{r^2}  - \xi {\cal R}\right\}\phi (\rho,\rho') = \fr
{\delta (\rho - \rho')}{r^2}
\ed
or in dimensionless variables ($\lambda\to i\nu k,\ k\rho \to x $)
\bd
\left\{\partial^2_x + 2 \fr {\dot{\rk}}\rk \partial_x - \nu^2 (1 +
\fr 1{\rk^2}) + (\fr  1{4\rk^2} - \xi
\mathcal{R}_k)\right\}\phi(x,x') = \fr {k\delta (x - x)}{\rk^2}.
\ed

As usual we represent the radial Green function in standard form:
\bd
\phi (x,x') = \theta (x'-x) \phi_1(x) \phi_2(x') + \theta (x-x')
\phi_2(x) \phi_1(x'),
\ed
where $\phi_1$ and $\phi_2$ are two linearly independent solutions
of homogenous equation and $\phi_1$ tends to infinity for $\rho\to
\infty$ and $\phi_2$ tends to $0$. The Wronskian condition appears
if we substitute this form of radial Green function to radial
equation above:
\bd
W (\phi_1(x),\phi_2(x')) = \fr {k}{\rk^2}.
\ed
Therefore, if two functions $\phi_1$ and $\phi_2$ describe the
system, they have to obey this Wronskian condition.

For solution in exponential form \Ref{ExpFor} this condition gives
\bd
e^{S^+(x) + S^-(x)} = \fr {2\nu \ak}{\rk^2} \fr  1{\dot{S}^+(x) -
\dot{S}^-(x)}.
\ed
The denominator in rhs has the following form
\bd
\dot{S}^+(x) - \dot{S}^-(x) = 2\sum_{n=0}^\infty \nu^{1-2n}
\dot{S}^+_{2n-1}.
\ed
Taking into account these two  expressions above we arrive at the
formula
\be\label{ME}
S^+(x) + S^-(-x) = \ln (\ak) - \fr  12 \ln (\dot{S}^2_{-1} \rk^4)
+ \sum_{n=0}^\infty \nu^{1-2n}\int_{-x}^{+x}\dot{S}^+_{2n-1}dx -
\ln \left\{1 + \sum_{n=1}^\infty \nu^{-2n}\fr
{\dot{S}^+_{2n-1}}{\dot{S}^+_{-1}}\right\}.
\ee

The main achievements and peculiarities of this expression are:
(i) the rhs is expressed in terms of derivative of functions
$S^+_n$, we do not need to find the constants of integration in
the chain of equations \Ref{System}, (ii) the odd and even power
of $\nu$ are separated, which leads to separation of the contribution to
heat kernel coefficients with integer and half-integer indices,
(iii) the rhs is expressed in terms of functions $S_n$ with odd
indices, only. The first three functions $\dot{S}_{2n-1}$ are
listed in Appendix \ref{A}, formula \Ref{SeriesS}. We would like
to note that this formula is valid for arbitrary, but symmetric,
$r(\rho) = r(-\rho)$, metric coefficient.

From this expression we may conclude that the contribution from
the second term in condition \Ref{MainCond} is exponentially small
comparing with first one. Indeed, the main WKB term in Eq.
\Ref{ME} gives the following contribution:
\bnn
\phi_1(+R)\phi_2(-R) \approx \fr {k}{2\nu}\fr 1{\dot{S}^+_{-1}
\rk^2}\exp \left\{+\nu \int_{-kR}^{+kR}\dot{S}^+_{-
1}dx\right\}, \adb\\
\phi_1(-R)\phi_2(+R) \approx \fr {k}{2\nu}\fr 1{\dot{S}^+_{-1}
\rk^2}\exp \left\{-\nu \int_{-kR}^{+kR}\dot{S}^+_{- 1}dx\right\}.
\enn

Because the function $\dot{S}^+_{-1}$ is positive for arbitrary
$R$ we observe that the second expression gives exponentially
small (for $R\to \infty$) contribution comparing with first one
and we will omit it in what follows.

\section{Heat kernel coefficients}\label{Heat}

Let us now proceed to evaluation of the heat kernel coefficients
(HKC). The formula \Ref{ME} allows us to find HKC in general form
for arbitrary indices. Taking into account above discussions we
have the following expression for zeta- function
\be \label{Zeta1}
\zeta(s - \fr  12) = - m^{2s}\fr {2\cos\pi s}\pi \sum_{l=0}^\infty
\nu^{2-2s} \int_{\fr  m\nu}^\infty dk (k^2 - \fr
{m^2}{\nu^2})^{1/2 -s} \fr {\partial}{\partial k} \left\{S^+(+R) +
S^-(-R)\right\}.
\ee

To find the heat kernel coefficients we use the uniform expansion
given by Eq. \Ref{ME}. As it will be clear later, the odd powers
of $\nu$ will give contribution to HKC with integer indices and
even powers of $\nu$ produce the contribution to HKC with
half-integer indices. The well-known asymptotic expansion of
zeta-function in three dimensions has the form below
\be\label{Zeta3Dim}
\zeta_{as} (s-\fr  12) = \fr {1}{(4\pi)^{3/2}}\fr  1{\Gamma (s-\fr
12)} \sum_{l=0}^\infty \left\{ m^{4-2l} B_l \Gamma (s+l-2) +
m^{3-2l} B_{l+1/2} \Gamma (s+l-\fr  32)\right\}.
\ee
For simplicity we introduce the density of HKC with integer
indices $\overline{B}_l$ by relation
\bd
B_l = \int_{-R}^{+R} d\rho \overline{B}_l(\rho)
\ed
and first of all we will obtain formulas for this density.

Let us consider the part of Eq. \Ref{ME} with odd degree of $\nu$.
The contribution to zeta-function is the following:
\bd
\zeta^{odd}_{as}(s - \fr  12) = - m^{2s}\fr {2\cos\pi s}\pi
\sum_{l=0}^\infty \nu^{2-2s} \int_{\fr  m\nu}^\infty dk (k^2 - \fr
{m^2}{\nu^2})^{1/2 -s} \fr {\partial}{\partial k}  \left\{
\sum_{p=0}^\infty \nu^{1-2p}\int_{-kR}^{+kR}\dot{S}^+_{2p-1}(x)dx
\right\}.
\ed
We change now the variable of integration $x=k\rho$ and take the
derivative with respect to $k$:
\bd
\zeta^{odd}_{as}(s - \fr  12) = - m^{2s}\fr {2\cos\pi s}\pi
\sum_{l=0}^\infty \nu^{2-2s} \int_{\fr  m\nu}^\infty dk k (k^2 -
\fr {m^2}{\nu^2})^{1/2 -s}  \sum_{p=0}^\infty
\nu^{1-2p}\int_{-R}^{+R}s_{2p-1}(k,\rho)d\rho.
\ed
The first four functions $s_{2p-1}$ are listed in Appendix
\ref{A}, formula \Ref{Sodd}. The general structure of these
functions is the following
\bd
s_{2p-1} = \sum_{n=0}^{2p} \alpha_{2p-1,n} z^{-p-n-\fr 12},
\ed
where $\alpha_{2p-1,n}$ are the functions of $r(\rho)$, and $z= 1
+ k^2 r^2(\rho)$.

Next, we integrate over $k$ using the formula
\bd
\int_{\fr  ma}^\infty dk k (k^2 - \fr  {m^2}{\nu^2})^{\fr  12 -s}
(1+k^2r^2)^{-q} = \fr  12 r^{2s-3}\nu^{-3+2s+2q} (\nu^2 +
m^2r^2)^{\fr  32 -s -q} \fr {\Gamma (\fr  32- s)\Gamma (q-\fr
32+s)}{\Gamma (q)}
\ed
and obtain the following expression for odd part of the zeta
function
\be\label{ZetaOdd}
\zeta^{odd}_{as}(s - \fr  12) = \fr {m^{2s}}{\Gamma (s - \fr  12)}
\int_{-R}^{+R} d\rho \sum_{l=0}^\infty \sum_{p=0}^\infty
\sum_{n=0}^{2p} \alpha_{2p-1,n} r^{2s-3} \fr {\Gamma
(s+p+n-1)}{\Gamma (p+n+\fr  12)} \fr {\nu^{2n+1}}{(\nu^2 +
m^2r^2)^{s+p+n- 1}}.
\ee

By using the binomial of Newton we reduce the power of $\nu$ in
nominator
\be\label{Binom}
\sum_{l=0}^\infty \fr {\nu^{2n+1}}{(\nu^2 + m^2r^2)^{s+p+n-1}} =
\fr  12 \sum_{q=0}^n (-m^2r^2)^{n-q} \fr {n!}{q!(n-q)!} \fr {{\cal
Z} (s+p+n-1-q)}{\Gamma (s+p+n-1-q)},
\ee
where
\bd
{\cal Z}(s) = 2\Gamma (s) \sum_{l=0}^\infty \fr  \nu{(\nu^2 +
m^2r^2)^s}.
\ed

To obtain the HKC we need asymptotic (over mass $m$) expansion
of the zeta-function. The asymptotic expansion of function ${\cal
Z}(s)$ was obtained in the Ref. \cite{BezBezKhu01} and it has the
form below
\bs\label{Z}
\bn
{\cal Z}(s) &=& (mr)^{-2s} \sum_{l=-1}^\infty A_l(s) (mr)^{-2l},\adb\\
A_{-1}(s) &=& \Gamma (s-1),\adb\\
A_l(s) &=& 2 \fr {(-1)^l}{l!} \Gamma (l+s) \zeta_H (-1 - 2l,\fr
12),
\en
\es
where the $\zeta_H(a,b)$ is the Hurwitz zeta-function.

Taking into account formulas \Ref{Binom} and \Ref{Z} one has the following
asymptotic series for odd part of zeta-function
\bnn
\zeta^{odd}_{as}(s - \fr  12) &=& \fr {1}{2\Gamma (s-\fr
12)}\int_{-R}^{+R}d\rho \sum_{l=0}^\infty \sum_{p=0}^l
\sum_{n=0}^{2p} \sum_{q=0}^{n}\alpha_{2p-1,n}
m^{4-2l}r^{-2l+1} \fr {\Gamma (p+n-1+s)}{\Gamma (p+n+\fr  12)}\adb\\
&\times& \fr {n!}{q!(n-q)!} (-1)^{n-q} \fr
{A_{l-p-1}(s+p+n-1-q)}{\Gamma (s+p+n-1- q)}.\nonumber
\enn

As it was expected at the beginning this is series over even
degree of mass and it gives contribution to the HKC with integer
indices. Comparing above expression with general \Ref{Zeta3Dim} we
obtain general formula for arbitrary HKC coefficient with integer
index
\be\label{hkcdensity}
\overline{B}_l(\rho) = \fr {4\pi^{3/2}}{\Gamma (s+l-2)}
\sum_{p=0}^l \sum_{n=0}^{2p}
\sum_{q=0}^{n}\alpha_{2p-1,n}r^{-2l+1} \fr {\Gamma (p+n-
1+s)}{\Gamma (p+n+\fr  12)}\fr {n! (-1)^{n-q}}{q!(n-q)!} \fr
{A_{l-p-1}(s+p+n-1- q)}{\Gamma (s+p+n-1-q)}.
\ee
Therefore, to obtain HKC with index $l$ we have to take into
account expansion up to $\nu^{1-2l}$.

Let us now proceed to HKC with half-integer indices. To find them
we have to take into account the rest part of Eq. \Ref{ME} with
even powers of $\nu$ in expression for zeta-function \Ref{Zeta1}.
The general form of even part is
\be\label{SEven}
(S^+(x) + S^-(-x))^{even} = \ln (\ak) - \fr  12 \ln
(\dot{S}^2_{-1} \rk^4) - \ln \left\{1 + \sum_{n=1}^\infty
\nu^{-2n}\fr {\dot{S}^+_{2n-1}}{\dot{S}^+_{- 1}}\right\} =
\sum_{p=0}^\infty \nu^{-2p} E_{2p},
\ee
where the first four functions $E_{2p}$ are listed in Appendix
\ref{A}, formula \Ref{SeriesE}.

We substitute now the expansion \Ref{SEven} in the expression for
zeta-function:
\bd
\zeta^{even}_{as}(s - \fr  12) = - m^{2s}\fr {2\cos\pi s}\pi
\sum_{l=0}^\infty \nu^{2-2s} \int_{\fr  m\nu}^\infty dk (k^2 - \fr
{m^2}{\nu^2})^{1/2 -s} \fr {\partial}{\partial k}
\sum_{p=0}^\infty \nu^{-2p}E_{2p},
\ed
and take the derivative with respect to the $k$:
\be\label{ZetaEven}
\zeta^{even}_{as}(s - \fr  12) = - m^{2s}\fr {2\cos\pi s}\pi
\sum_{l=0}^\infty \nu^{2-2s} \int_{\fr  m\nu}^\infty dk k (k^2 -
\fr {m^2}{\nu^2})^{1/2 -s}  \sum_{p=0}^\infty \nu^{-2p}s_{2p}.
\ee

The functions $s_{2p}$ have the following structure
\bd
s_{2p} = \sum_{n=0}^{2p} \alpha _{2p,n} z^{-p-n -1},
\ed
where $z = 1 + k^2 r^2(R)$. The first three coefficients are
listed in manifest form in Appendix A (see Eq. \Ref{Seven}). The
coefficients $\alpha _{2p,n}$ depend on the parameters $R$ and $a$
and do not depend on the variable of integration $k$. Going
the same way as we did for HKC with integer indices we obtain the
following asymptotic expression for even part of the zeta-function:
\bnn
\zeta^{even}_{as}(s - \fr  12) &=& \fr {1}{2\Gamma (s-\fr  12)}
\sum_{l=0}^\infty \sum_{p=0}^l \sum_{n=0}^{2p}
\sum_{q=0}^{n}\alpha _{2p,n}
m^{3-2l}r^{-2l} \fr {\Gamma (p+n-\fr  12+s)}{\Gamma (p+n+1)}\adb\\
&\times& \fr {n!}{q!(n-q)!} (-1)^{n-q} \fr {A_{l-p-1}(s+p+n-\fr
12-q)}{\Gamma (s+p+n -\fr  12-q)}\nonumber
\enn

As it was expected the even part of zeta-function is the series over odd
powers of mass and, therefore, it gives contributions to HKC
with half-integer indices. Comparing this expression with general
asymptotic series for zeta-function we obtain the following
formula for HKC with half-integer indices:
\be\label{hkchalf}
B_{l+\fr  12} = \fr {4\pi^{3/2}}{\Gamma (s+l-\fr  32)}
\sum_{p=0}^l \sum_{n=0}^{2p} \sum_{q=0}^{n}\alpha _{2p,n}r^{-2l}
\fr {\Gamma (p+n- \fr  12 + s)}{\Gamma (p+n+1)}\fr {n!
(-1)^{n-q}}{q!(n-q)!} \fr {A_{l-p-1}(s+p+n - \fr  12 - q)}{\Gamma
(s+p+n-\fr  12-q)}.
\ee
We would like to note that the right hand side of formulas
\Ref{hkcdensity} and \Ref{hkchalf} does not depend, in fact, on
the $s$, which is confirmed by straightforward calculations.

These formulas look very complicate but calculation may be done
easily using simple program in package {\it Mathematica}. Indeed,
the functions $\dot{S}^+_2(x)$ and $E_{2n}$ may be found by using
formulas \Ref{System} and \Ref{SEven}. The functions $s_n$ are
obtained from the following relations:
\bnn
s_{2n-1}(k,\rho) &=& \fr 1k \fr\partial{\partial k}\left[ \left.k
\dot{S}^+_{2n-1}(x)\right|_{x=k\rho} \right],\\
s_{2n}(k,R) &=& \fr 1k \fr\partial{\partial k}\left[
\left.E_{2n}(x)\right|_{x=kR} \right].
\enn

The first four HKC coefficient (density) with integer indices are
listed below
\bs\label{hkcdensitygene}
\bn
\overline{B}_0 &=& 4\pi{r}^2,\adb\\
\overline{B}_1 &=& \frac{8\pi}{3} \left[{r'}^2 + rr''\right] +
8\pi \left( \xi - \frac{1}{6}\right) \left[ -1 + {r'}^2 + 2rr''\right],\adb\label{OverB1}\\
\overline{B}_2 &=& \frac{8\pi {\xi}^2}{{r}^2}\left[ -1 + {r'}^2 +
2rr''\right]^2 + \frac{8\pi \xi}{3{r}^2}\left[\left( -1 + r'^2
\right)  - r\left( -5 + 7 r'^2 \right) r'' +
7r^2r'r^{(3)} + 3r^3r^{(4)}\right]\nonumber\adb\\
&-& \frac{2\pi}{315{r}^2}\left[ 2\left( -21 + 17r'^4 \right) -
6r\left( -35 + 59r'^2 \right) r'' + 21r^2\left( 7r''^2 +
24r'r^{(3)} \right)  + 210r^3r^{(4)} \right],\adb\label{OverB2}\\
\overline{B}_3 &=& \frac{16\pi{\xi}^3}{3{r}^4}\left[ -1 + {r'}^2 +
2rr''\right]^3 + \frac{8\pi{\xi}^2}{3{r}^4}\left[ \left( -1 + r'^2
\right)^2\left( 1 + 9r'^2 \right) - 2r\left( -1 + r'^2 \right)
\left( -5 + 9r'^2 \right) r''\nonumber\right.\adb\\
&-&\left. 2r^2\left( -8r''^2 + 16r'^2r''^2 - 3r'r^{(3)} +
3r'^3r^{(3)} \right) + 2r^3\left( 14r'r''r^{(3)} - 3r^{(4)} +
3r'^2r^{(4)}\right) + 2r^4\left( 5r^{(3)}{}^2 + 6r''r^{(4)}
\right) \right] \nonumber\adb\\
&-& \frac{4\pi\xi}{315{r}^4}\left[-42\left( -1 + r'^2\right)
\left( 1 + 15r'^2 \right)  +  2r\left( -252 + 105r'^2 + 859r'^4
\right)r''\nonumber\right.\adb\\
&-&\left. 2r^2\left( -525r''^2 + 2517r'^2r''^2 - 420r'r^{(3)} +
808r'^3r^{(3)}\right) + 3r^3 \left( 308r''^3 + 1354r'r''r^{(3)} -
175r^{(4)} + 271r'^2r^{(4)}\right)\nonumber\right.\adb\\
&+&\left. 21r^4\left( 27r^{(3)}{}^2 + 27r''r^{(4)} -
13r'r^{(5)}\right) - 105r^5r^{(6)}\right] - \frac{\pi}{45045{r}^4}
\left[4\left( -572 - 9009r'^2 + 9341r'^6
\right)\nonumber\right.\adb\\
&-&\left. 4r\left( -6006 - 15015r'^2 + 62039r'^4 \right) r'' +
13r^2\left(-4620r''^2 + 32943r'^2r''^2 - 4620r'r^{(3)} +
11564r'^3r^{(3)}\right)\right.\adb\label{OverB3}\\
&-&\left. 286r^3 \left( 308r''^3 + 1139r'r''r^{(3)} - 105r^{(4)} +
223r'^2r^{(4)}\right) - 429r^4\left( 47r^{(3)}{}^2 + 24r''r^{(4)}
- 74r'r^{(5)}\right)  + 12012r^5r^{(6)} \right].\nonumber
\en
\es
In above formulas the function $r$ depends on the radial
coordinate $\rho$ whereas the heat kernel coefficients with
half-integer indices
\bs\label{hkchalfgen}
\bn
B_{1/2} &=& -4\pi^{3/2}r^2,\adb\\
B_{3/2} &=& -8\pi^{3/2}\xi \left[ -1 + r'^2 + 2rr''\right] +
\frac{\pi^{3/2}}{3}\left[ -4 + 3r'^2 + 6rr''\right],\adb\\
B_{5/2} &=& \frac{-8\pi^{3/2}\xi^2}{r^2} \left[ -1 + r'^2 +
2rr''\right]^2 - \frac{2\pi^{3/2}\xi}{3r^2} \left[4\left( -1 +
r'^2\right)  - 10r\left( -2 + 3r'^2 \right) r'' - 3r^2\left(
4r''^2 - 3r'r^{(3)}\right) + 6r^3r^{(4)}\right] \adb\nonumber\\
&+& \frac{\pi^{3/2}}{120r^2} \left[2\left( -16 + 15r'^4 \right) -
5r\left( -32 + 63r'^2 \right) r'' - 10r^2\left( 5r''^2 -
14r'r^{(3)}\right)  + 90r^3r^{(4)}\right],
\en
\es
depend on the radial function $r$ at boundary: $r=r(R)$. From Eqs.
\Ref{hkcdensitygene} and \Ref{hkchalfgen}  we observe that the HKC
$B_l$ and $B_{l+1/2}$ are polynomial in $\xi$ with degree $l$.

It is well-known \cite{ZetaBook} that the heat kernel coefficients
with integer indices consist of two parts. First part is an
integral over volume and another one is an integral over boundary.
We obtained slightly different representation for this coefficient
as integral over $\rho$. But it is easy to see that they are in
agreement. Indeed, let us consider for example coefficient $B_1$.
According with standard formula we have
\bd
B_1 = \left(\fr 16 -\xi\right) \int_V \mathcal{R} dV + \left.\fr
13 \int_S tr K dS\right|\strut_{\rho=+R} + \left.\fr 13 \int_S tr
K dS\right|\strut_{\rho=-R}.
\ed
Volume contribution is exactly the same as we have already
obtained \Ref{OverB1}. Surface contribution from above formula is
\bd
\fr 13 \int_S tr K dS\strut_{\rho=+R} + \fr 13 \int_S tr K
dS\strut_{\rho=-R} = \left.\fr{16\pi}{3}r'r\right|\strut_{\rho=R}.
\ed
From our result \Ref{OverB1} we get the same expression
\bd
\frac{8\pi}{3}\int_{-R}^{+R} \left[{r'}^2 + rr''\right]d\rho =
\frac{8\pi}{3}\int_{-R}^{+R} [rr']'d\rho = \left.\frac{16\pi}{3}
rr'\right|\strut_{\rho=R}.
\ed
It is not so difficult to verify that the heat kernel coefficients
up to $B_2$ are in agreement with general expressions. There is no
general expressions for higher coefficients.

According with Ref. \cite{KhuSus02} the sufficient condition for
existing of the the self-consistent wormholes may be formulated in
terms of two heat kernel coefficients
\bnn
B_2 &=& \int_{-\infty}^{+\infty} d\rho \overline{B}_2 =
h_{2,2}\xi^2 + h_{2,1}\xi + h_{2,0},\adb\\
B_3 &=& \int_{-\infty}^{+\infty} d\rho \overline{B}_3 =
h_{3,3}\xi^3 + h_{3,2}\xi^2 + h_{3,1}\xi + h_{3,0}.
\enn
Namely, both $B_2$ and $B_3$ have to be positive \footnote{We
would like to note that in Ref. \cite{KhuSus02} instead of $B_3$
the coefficient $B_{5/2}$ appears. It is connected with specific
form of wormholes. The point is that the background contains
singular surface of codimension one.}. The coefficients $h_{k,l}$
of polynomials depend on the structure of wormhole. Therefore the
problem reduces to analysis of polynomial in $\xi$ of second and
third degree, the coefficients of which depend on the structure of
wormhole's space-time. Wormholes with different forms may exist
for different values of non-minimal connection $\xi$, or vice
versa for some $\xi$ the above polynomials will be positive for
specific form of wormholes.

\section{The model of throat: $r(\rho) = \sqrt{\rho^2 +
a^2}$}\label{Model1}

In this section we consider in detail the specific model of
wormhole with the following profile of throat $r(\rho) =
\sqrt{\rho^2 + a^2}$. From general expressions
\Ref{hkcdensitygene} we obtain the density of heat kernel
coefficients with integer indices which are
\bnn
\overline{B}_{0} &=& 4\pi r^2,\adb\\
\overline{B}_{1} &=& \fr {8\pi a^2}{r^2} (\xi - \fr  16) + \fr {8\pi}3,\adb\\
\overline{B}_{2} &=& \frac{2\pi }{315r^6}\left( 1103a^4 -
796a^2r^2 + 8r^4 \right) - \frac{8a^2\pi}{3r^6} \left( 17a^2 -
12r^2 \right)\xi
+ \fr {8\pi a^4}{r^5}\xi^2,\adb\\
\overline{B}_{3} &=& -\frac{2\pi}{45045r^{10}}\left( 2583561a^6 -
3157438a^4r^2 + 751820a^2r^4 - 480r^6 \right) +
\frac{4a^2\pi }{315r^{10}}\left( 47263a^4 - 57464a^2r^2 + 13540r^4 \right)\xi \adb\\
&-& \frac{8a^4\pi}{3r^{10}} \left( 73a^2 - 62r^2 \right) \xi^2 +
\fr {16\pi a^6}{3 r^{10}}\xi^3\nonumber
\enn
Integrating over $\rho$ from $-R$ to $+R$ we obtain HKC. Below we
reproduce their expansions in the limit $R\to\infty$ up to terms
$1/R$:
\bs\label{HKCRtoInfty}
\bn
B_0 &=& \fr {8\pi R^3}3 + 8\pi a^2 R,\adb\\
B_1 &\approx&  \fr {16\pi R}3 + 8\pi^2 a
(\xi - \fr  16) - \fr {16\pi a^2}R (\xi - \fr 16),\adb\\
B_2 &\approx&  \fr {\pi^2}{20 a} (60\xi^2 - 20\xi +3)
- \fr {32\pi}{315 R},\label{B2}\adb\\
B_3 &\approx&  \fr {\pi^2}{4032 a^3} (5880\xi^3 - 6300\xi^2 +
2226\xi -257). \label{B3}
\en
\es
The formulas for first three coefficients with half integer
indices may be found from general expression \Ref{hkchalfgen}.
Below we have listed them with their expansions for large value of
$R$
\bnn
B_{\fr  12} &=& -4\pi^{3/2} r^2= -4\pi^{3/2}(R^2 + a^2),\adb\\
B_{\fr  32} &\approx& -\fr {\pi^{3/2}}3 -\fr {\pi^{3/2}a^2 (8\xi
-1)}{60R^2},\adb\\
B_{\fr  52} &\approx&  -\fr {\pi^{3/2}}{60R^2}.
\enn

Let us now proceed to renormalization and calculation of the zero
point energy. As noted in Sec. \ref{ZeroPoint} (see Eq.
\Ref{GenForm}) we have to subtract all terms which will survive in
the limit $m\to \infty$. According with general asymptotic
structure of the zeta-function given by Eq. \Ref{Zeta3Dim} in this
limit the first five terms survive, namely HKC up to $B_{2}$.
Because the zero point energy is proportional to zeta-function we
may speak about renormalization of the zeta function. According with
Eq. \Ref{GenForm} we take asymptotic expansion for zeta-function
up to $\nu^{-3}$ (in the limit $m\to \infty$ these terms give
asymptotic (over $m$) expansion \Ref{Zeta3Dim} up to heat kernel
coefficient $B_2$) and subtract its expansion over $m$ up
to $B_2$ from it. After taking the limit $s\to 0$ we observe that
this difference will
give $E_{as}^{fin}$ \Ref{Easfin}. First of all we consider this
part and later will simplify the finite part \Ref{Efin}.

We should like to make a comment. In the problem under
consideration we have two different scales: $R$ and $a$ which give
us two dimensionless parameters $mR$ and $ma$. To extract terms
for renormalization we turn mass to infinity, which means the
Compone wavelength of scalar boson turns to zero and becomes
smaller then all scales of model. In other words it means that we
turn to infinity both parameters $mR$ and $ma$. After
renormalization we will turn $mR$ to infinity separately in order
to obtain the part which does not depend on the boundary.

Let us consider separately two parts of asymptotic expansion of
zeta-function according with odd and even powers of $\nu$. First
of all we consider odd part which gives the HKC with integer
indices. All singularities are contained in the first three terms
in Ed. \Ref{Zeta3Dim} with $B_0,\ B_1,\ B_2$. After subtracting
these singularities we tend $s\to 0$ and obtain some infinite
power series over parameters $m\rho$ and $ma$. Next, we have to
integrate over $\rho$ and tend $mR\to\infty$. For this reason we
have to obtain some expression instead of series to take  this
limit easily. It is impossible to take this limit directly in series. We
will use the Abel-Plana formula to extract the main
contribution from series in this limit. The rest will be a
good expression for
numerical calculations. Moreover, from this rest part we will
extract terms which will be divergent in the limit $ma \to 0$ to
analyze our formulas.

Our starting formula for odd part of zeta-function is Eq.\Ref{ZetaOdd}
which cut up to $p=2$:
\be\label{ZetaOdd1}
\zeta^{odd}_{as,2}(s - \fr  12) = \fr {m^{2s}}{\Gamma (s - \fr
12)} \int_{-R}^{+R} d\rho \sum_{l=0}^\infty \sum_{p=0}^2
\sum_{k=0}^{2p} \alpha_{2p-1,k} r^{2s-3} \fr {\Gamma
(s+p+k-1)}{\Gamma (p+k+\fr 12)} \fr {\nu^{2k+1}}{(\nu^2 +
m^2r^2)^{s+p+k-1}}.
\ee
Expanding the denominator with the help of the formula
\bd
(1+x^2)^{-s} = \sum_{n=0}^\infty \fr {(-1)^n}{n!} \fr {\Gamma
(n+s)}{\Gamma (s)} x^{2n},
\ed
we represent  Eq. \Ref{ZetaOdd1} in the following form
\be\label{Odd}
\zeta^{odd}_{as,2}(s - \fr  12) = \fr {1}{(4\pi)^{3/2}}\fr
{1}{\Gamma (s - \fr  12)} \int_{- R}^{+R} d\rho m^{2s}
\sum_{n=0}^\infty\sum_{p=0}^2 m^{2n} f_{n,p}(s),
\ee
where
\bd
f_{n,p}(s) = 8\pi^{3/2} \fr {(-1)^n}{n!} r^{2n-3} \sum_{k=0}^{2p}
\alpha_{2p- 1,k}\fr {\Gamma (n+p+k+s-1)}{\Gamma (p+k+\fr  12)}
\zeta_H (2n+2p+2s-3,\fr  12).
\ed

In order to make formulas more readable we make everything dimensionless but
save the same notations. In any moment we may repair dimensional parameters by
changing $R\to mR$ an $a\to ma$. In this case we rewrite the expression for
zeta-function \Ref{ZetaOdd1} in the following form
\be\label{Oddd}
\zeta^{odd}_{as,2}(s - \fr  12) = \fr {m}{(4\pi)^{3/2}}\fr {
m^{2s}}{\Gamma (s - \fr  12)} \int_{-R}^{+R} d\rho
\sum_{n=0}^\infty\sum_{p=0}^2 f_{n,p}(s),
\ee

From this expression we observe that for $p=0$ the first three terms are
divergent with $n=0,1,2$; for $p=1$ the first two terms with $n=0,1$ and at last
for $p=2$ the only term is divergent with $n=0$. We remind that
\bd
\Gamma (s-n)\strut_{s\to 0} = \fr {(-1)^n}{n!} \left(\fr  1s +
\Psi (n+1)\right) + O(s),\ \zeta_H (s+1,q)\strut _{s\to 0} = \fr
1s - \Psi (q) + O(s).
\ed
For this reason we represent zeta-function \Ref{ZetaOdd1} in the
following form (for $s\to 0$)
\bs\label{OddGen}
\bn
\zeta^{odd}_{as,2}(s - \fr  12) &=& \fr {m}{(4\pi)^{3/2}}\fr
{1}{\Gamma (s - \fr  12)} \int_{-R}^{+R} d\rho \left\{
r^{2s}\sum_{n=0}^2 f_{n,0}(s) + \sum_{n=3}^\infty
f_{n,0}(0)\right\}\label{OddGen1}\adb\\
&+& \fr {m}{(4\pi)^{3/2}}\fr {1}{\Gamma (s - \fr  12)}
\int_{-R}^{+R} d\rho \left\{ r^{2s}\sum_{n=0}^1 f_{n,1}(s) +
\sum_{n=2}^\infty
f_{n,1}(0)\right\}\label{OddGen2}\adb\\
&+& \fr {m}{(4\pi)^{3/2}}\fr {1}{\Gamma (s - \fr  12)}
\int_{-R}^{+R} d\rho \left\{ r^{2s}f_{0,2}(s) + \sum_{n=1}^\infty
f_{n,2}(0)\right\}\label{OddGen3}
\en
\es
and will analyze each part separately.

To illustrate the calculations  we consider in details the first
part \Ref{OddGen1}. First of all it is not difficult to find the
manifest form of singular part in the limit $s\to 0$:
\bnn
\sum_{n=0}^2 f_{n,0} (s) &=& 4\pi r^2 \Gamma (s-2) + \fr {\pi}3
\Gamma (s-1) +
\fr {7\pi}{120r^2} \Gamma (s) + \pi r^2 [-3 + 4\gamma + 8 \ln (2)]\adb\\
&+& \fr {\pi}3 [1+2\ln (2) + 24 \zeta'_R (-1)] - \fr
{\pi}{120r^2}[-7 + 2\ln (2) - 1680 \zeta'_R (-3)]. \nonumber
\enn
We observe that this term gives contribution to $B_0,\ B_1,$ and
$B_2$ according with gamma functions. For renormalization we have
to subtract from this expression the first three terms according
to our scheme.

There is one important moment which is crucial for our analysis.
Above formula contains all terms which survive in the limit $s\to
0$ for arbitrary mass of field. For renormalization we have to
subtract asymptotic expansion in the form \Ref{Zeta3Dim}. There is a
difference in factor $r^{2s}$. For this reason after
renormalization factor
\bd
(r^{2s} - 1) (4\pi r^2 \Gamma (s-2) + \fr {\pi}3 \Gamma (s-1) +
\fr {7\pi}{120r^2} \Gamma (s))\strut_{s\to 0} = \left[2\pi r^2  -
\fr {\pi}3 + \fr {7\pi}{120r^2} \right] \ln r^2
\ed
appears. If we take into account all terms in Eq. \Ref{OddGen} we
obtain the following contribution
\be\label{LogTerms}
\ln (r^2) (\fr  12 \overline{B}_0 - \overline{B}_1
+\overline{B}_2).
\ee
Exactly the same structure was observed before in Refs.
\cite{BezBezKhu01,KhuSus02}. This term defines the behavior of
energy for small size of wormhole because it is maximally
divergent for small size of wormhole.

Therefore the renormalized contribution is
\bnn
\sum_{n=0}^2 f_{n,0}^{ren} (s) &=& \left[2\pi r^2  - \fr {\pi}3  +
\fr {7\pi}{120r^2} \right] \ln r^2 + \pi r^2 [-3 + 4\gamma + 8 \ln (2)]\adb\\
&+& \fr {\pi}3 [1+2\ln (2) + 24 \zeta'_R (-1)] - \fr
{\pi}{120r^2}[-7 + 2\ln (2) - 1680 \zeta'_R (-3)]. \nonumber
\enn
We represent the finite part in the following form
\bd
\sum_{n=3}^\infty f_{n,0} (0) = \fr {8\pi}{r^2} \sum_{l=0}^\infty
\nu^3 \left\{\ln \left(1 + \fr {r^2}{\nu^2}\right) - \fr
{r^2}{\nu^2} + \fr  12 \fr {r^4}{\nu^4} + \fr
{r^2}{\nu^2}\left[\ln \left(1 + \fr {r^2}{\nu^2}\right) - \fr
{r^2}{\nu^2}\right]\right\}
\ed
by using standard series representation for Hurwitz zeta-function.
To find more suitable form for these series we use the Abel-Plana
formula and obtain
\bnn
\sum_{l=0}^\infty \nu^2 \left[\ln \left(1 + \fr
{r^2}{\nu^2}\right) - \fr {r^2}{\nu^2}\right]&=& -\fr  12 r^2 \ln
(r^2) - r^2 [2\ln (2) + \gamma - \fr  12] + 2 \int_0^\infty \fr
{d\nu\nu }{e^{2\pi\nu} + 1} \ln\left|1-
\fr {r^2}{\nu^2}\right|,\adb\\
\sum_{l=0}^\infty \nu^3 \left[\ln \left(1 + \fr
{r^2}{\nu^2}\right) - \fr {r^2}{\nu^2} + \fr  12 \fr
{r^4}{\nu^4}\right]&=& \fr  14 r^4 \ln (r^2) -\fr  1{24} r^2 + r^4
[2\ln (2) + \gamma - \fr  18] - 2 \int_0^\infty \fr {d\nu\nu^3
}{e^{2\pi\nu} + 1} \ln\left|1-\fr {r^2}{\nu^2}\right|.
\enn

Taking into account these formulas we have
\bd
\sum_{n=0}^2 f_{n,0} (0) + \sum_{n=3}^\infty f_{n,0}^{ren}(0) =
16\pi \int_0^\infty \fr {d\nu\nu^3 }{e^{2\pi\nu} + 1}\left\{
\ln\left|1-\fr {\nu^2}{r^2}\right| + \fr {\nu^2}{r^2}- \fr
{\nu^2}{r^2} \ln\left|1-\fr {\nu^2}{r^2}\right|\right\}.
\ed

We now integrate this formula over $\rho$ from $-R$ to $+R$
according with Eq. \Ref{OddGen1} and take the limit $R\to\infty$.
After this we arrive at the expression
\bnn
\Ref{OddGen1}&=&-\fr  m{16\pi^2}f_a=-\fr  m{16\pi^2}\left\{
f_a^{sing} +\omega_a\right\},\ {\rm where} \adb\\
f_a^{sing}&=&\fr  {7\pi^2}{60 a}\ln (a) + \fr  {\pi^2} a\left(\fr
7{120} + \fr  1{10} \ln (2) + 14 \zeta'_R (-3)\right).
\enn
The manifest form of the regular contribution has written out in
Appendix \ref{A} (see Eq. \ref{Omegaabc}). We extracted all terms
with logarithm and that which is singular for $a\to 0$ and
collected them in $f_a^{sing}$. The rest part, $\omega_a$, is a
regular contribution.

Using the same procedure for second and third parts we obtain the
following expressions
\bnn
\Ref{OddGen2}&=&-\fr  m{16\pi^2}f_b=-\fr  m{16\pi^2}\left\{
f_b^{sing} +\omega_b\right\},\ {\rm where} \adb\\
f_b^{sing} &=& \pi^2 \left[-16 a \left(\xi - \fr  16\right) + \fr
1a \left(\fr  23\xi - \fr  16\right)\right] \ln (a) + \fr  1a
\left[-\fr  13 (1 + 24\zeta'_R(-1))\xi + \fr  19 (1 + 18\zeta'_R
(-1)) \right]. \adb\\
\Ref{OddGen3}&=&-\fr  m{16\pi^2}f_c=-\fr  m{16\pi^2}\left\{
f_c^{sing} +\omega_c\right\},\ {\rm where} \adb\\
f_c^{sing} &=& \fr {\pi^2}a \left[6\xi^2 - \fr  83 \xi + \fr  7{20}\right] \ln (a) \adb\\
&+& \fr {\pi^2}a \left[\fr  12 (-7+12\gamma + 36 \ln 2)\xi^2 + \fr
16 (15 -16\gamma - 48\ln 2)\xi + \fr  1{360}(-143 + 126\gamma +
378\ln 2)\right]\nonumber
\enn
The manifest form of the regular contributions $\omega_b$ and
$\omega_c$ have written out in Appendix \ref{A} (Eq.
\Ref{Omegaabc}).

\begin{figure}
\centerline{\epsfxsize=8truecm\epsfbox{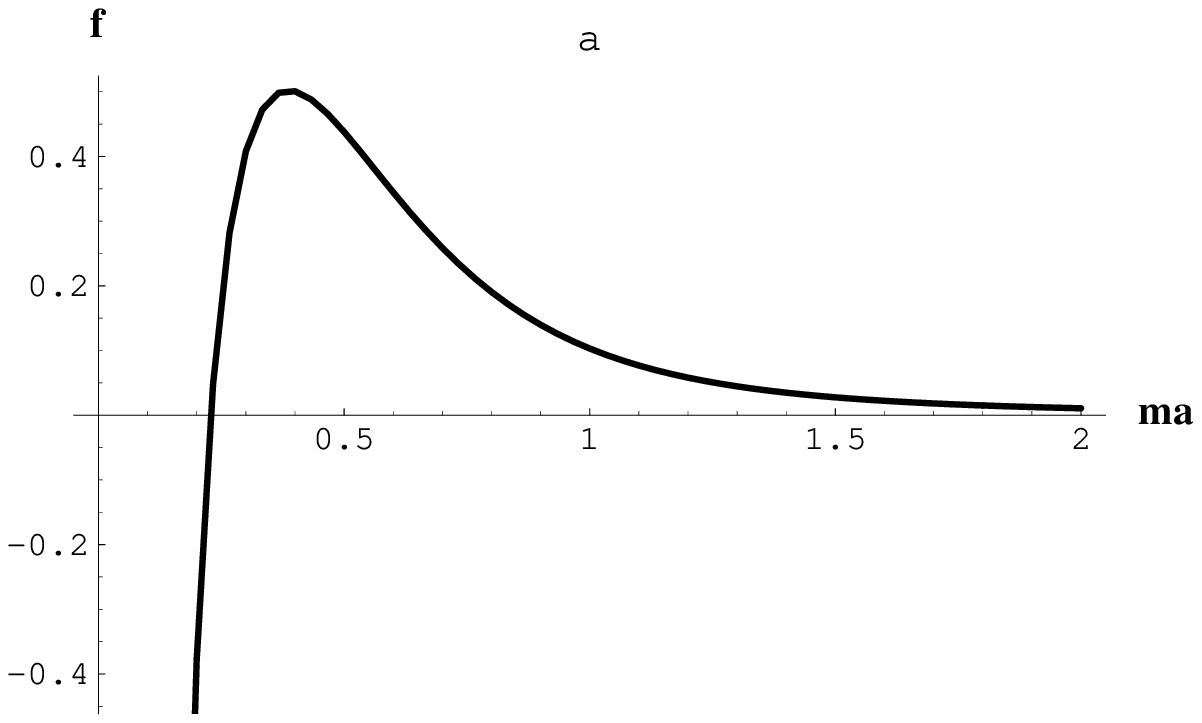}%
\epsfxsize=8truecm\epsfbox{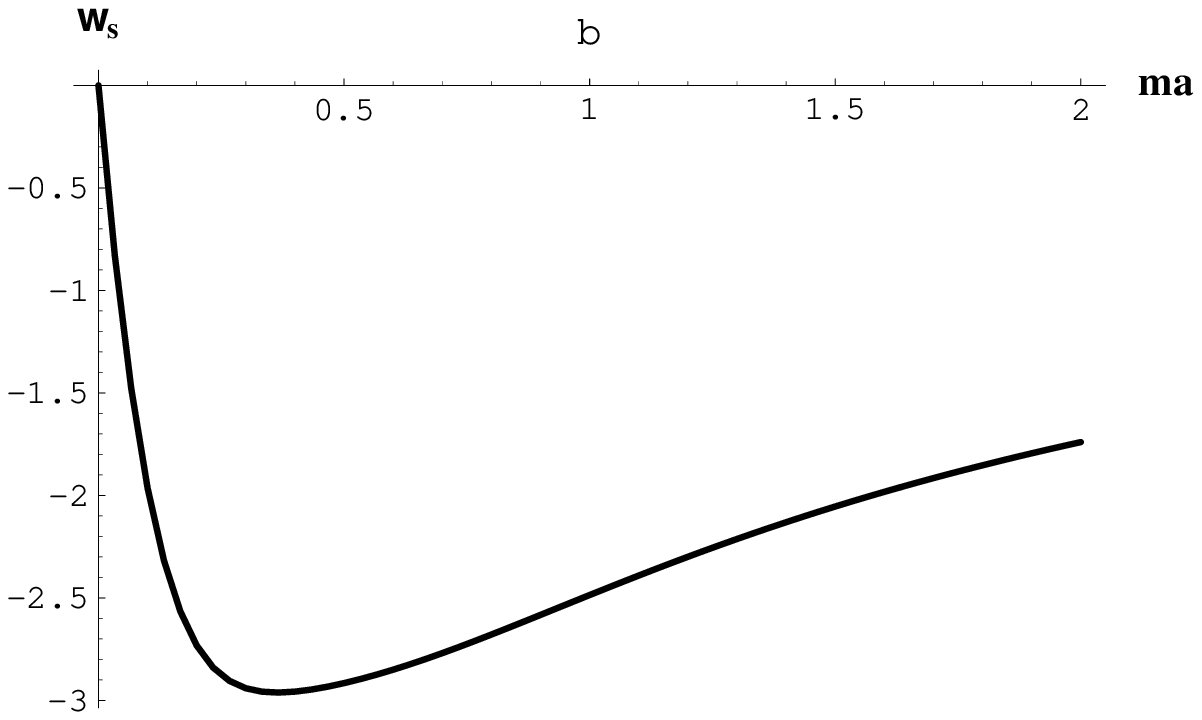}} \caption{The plot of the
summary contributions: $f=f_a + f_b + f_c$ and $\omega_s =
\omega_a + \omega_b + \omega_c$ for $\xi = \fr  16$: a) summary
contribution $f$ and b) regular part $\omega_s$} \label{fsum16}
\end{figure}

Putting together all contributions in \Ref{OddGen} we obtain
\bd
\zeta^{ren}_{odd} = -\fr  m{16\pi^2} \left\{\ln (a^2)
\pi^2\left(\fr  1a \left[3\xi^2 - 2\xi + \fr  3{20}\right] + 8
a\left(\xi - \fr  16 \right)\right) + \omega \right\},
\ed
where
\bnn
\omega &=& \omega_a + \omega_b + \omega_c + \fr {\pi^2}a \left[\fr
12(-7+12\gamma + 36\ln 2)\xi^2 - \fr  16 (-13+48\zeta'_R(-
1)+16\gamma + 48\ln 2)\right.\adb\\
&+&\left. \fr  1{180}(-41+360\zeta'_R(-1)+2520\zeta'_R(-3) +
63\gamma + 207\ln 2)\right]\nonumber
\enn
In Fig. \ref{fsum16} we reproduce a plot of sum of all three
contributions: $f = f_a + f_b + f_c$, $\omega_s = \omega_a +
\omega_b + \omega_c$ for $\xi = \fr 16$.

Let us now proceed to the contribution from even part of
zeta-function. We start from Eq. \Ref{ZetaEven} and do not take
the limit of great mass. Integrating over $k$ we obtain the
following expression for this even part
\be\label{ZetaEven1}
\zeta^{even}_{as}(s - \fr  12) = \fr {m^{2s}}{\Gamma (s - \fr 12)}
\sum_{l=0}^\infty \sum_{p=0}^1 \sum_{k=0}^{2p} \alpha _{2p,k}
r^{2s-3} \fr {\Gamma (s+p+k-\fr  12)}{\Gamma (p+k+ 1)} \fr
{\nu^{2k+1}}{(\nu^2 + m^2r^2)^{s+p+k- \fr  12}}.
\ee
Here $r$ is taken at the point $R$. Now we pass to dimensionless variables (with
the same notations) and  have
\be\label{ZetaEven2}
\zeta^{even}_{as}(s - \fr  12) = \fr {m}{(4\pi)^{3/2}\Gamma (s -
\fr 12)} \sum_{p=0}^1 \sum_{k=0}^{2p} \alpha _{2p,k} \fr
{(4\pi)^{3/2}}{r^3} \fr {\Gamma (s+p+k-\fr  12)}{\Gamma (p+k+ 1)}
\phi_{p,k},
\ee
where
\bd
\phi_{p,k}=\sum_{l=0}^\infty \fr {\nu^{2k+1}}{(\nu^2 +
m^2r^2)^{s+p+k- \fr  12}}
\ed

Analytical continuation $s\to 0$ in this series may be easily done
by using the Abel-Plana formula:
\bnn
\phi_{0,0} &=& -\fr  13 r^3  + \fr  1{24} r  - 2 r \int_r^\infty
d\nu \nu Ex(\nu) -
\fr  2r \int_0^r \fr {\nu^3 d\nu}{\sqrt{1-\fr {\nu^2}{r^2}}} Ex(\nu),\adb\\
\phi_{1,0} &=& - r + \fr  2r \int_0^r d\nu \nu [3 Ex(\nu) + \nu
Ex'(\nu)] \sqrt{1- \fr {\nu^2}{r^2}}\adb\\
\phi_{1,1} &=& -2 r + \fr  2r \int_0^r d\nu \nu [6 Ex(\nu) + 6\nu
Ex'(\nu) + \nu^2 Ex''(\nu)] \sqrt{1-\fr {\nu^2}{r^2}}\adb\\
\phi_{1,2} &=& -\fr  53 r + \fr  2r \int_0^r d\nu \nu [8 Ex(\nu) +
12 \nu Ex'(\nu) + 4 \nu^2 Ex''(\nu) + \fr  13 \nu^3 Ex'''(\nu)] \sqrt{1-\fr {\nu^2}{r^2}},\adb\\
\enn
where
\bd
Ex(\nu) = \fr  1{e^{2\pi \nu} + 1}.
\ed

Now we substitute these formulas in Eq. \Ref{ZetaEven2} and take
the limit $R\to \infty$. In this limit all integrals in
expressions for $\phi_{p,k}$ are smaller the $1/r$. Taking into
account that $\alpha _{0,0} \sim r$ and $\alpha _{1,k} \sim r^2$
we obtain in this limit (we repair dimensional variables)
\bd
\zeta^{even}_{as}(s - \fr  12) = \fr {m}{(4\pi)^{3/2}\Gamma (s -
\fr 12)} \left\{ m^3 B_{\fr  12} \Gamma (s - \fr  32) + m B_{\fr
32} \Gamma (s - \fr  12)  + O(\fr  1{R + a})\right\}.
\ed

Therefore after renormalization (subtracting these two terms) we take
the limit
$R\to \infty$ and obtain zero contribution from this even part
\bd
\zeta^{even}_{ren}(s - \fr  12) = 0.
\ed

Thus there is the only contribution from the odd part. Collecting
all terms together we arrive at the following expression for zero
point energy
\be\label{Erenfin}
E^{ren} = -\fr  m{32\pi^2} \left\{\ln (ma)^2 \pi^2\left(\fr 1{ma}
\left[3\xi^2 - \xi + \fr  3{20}\right] + 8ma\left(\xi - \fr  16
\right)\right) + \Omega \right\},
\ee
where
\bn
\Omega &=& \omega +  32\pi \sum_{l=0}^\infty \nu^{2}
\int_{\fr  1\nu}^\infty d y (y^2 - \fr {1}{\nu^2})^{1/2} \adb\\
&\times&\fr {\partial}{\partial y} \left\{S^+(+mR) + S^-(-mR) -
\left[S^+(+mR) + S^-(-mR)\right]_3^{uni.exp.} \right\}_{R\to
\infty}, \nonumber\label{Omega}
\en
and $k=my$.

The main problem now is the calculation of the last term in
expression for $\Omega$. Let us simplify the expression and show
that in the limit $R\to \infty$ the divergent parts are cancelled.
Indeed, let us consider the first five terms of uniform expansion
\bd
\left[S^+(R) + S^-(-R)\right]_3^{uni.exp.} = \sum_{n=0}^2
\nu^{1-2n}\int_{-kR}^{+kR}\dot{S}^+_{2n-1}dx + \sum_{n=0}^1
\nu^{-2n} E_{2n} .
\ed
It is very easy to take the limit in the part with even power of
$\nu$ by using the manifest form of $E_{2n}$ listed in Appendix
\ref{A}. The only term which gives the non-zero contribution is
 $E_0$:
\bnn
E_0 |_{R\to \infty} &=& -2 \ln (kR) + \ln (ka) + O(R^{-2}),\adb\\
E_2 |_{R\to \infty} &=&  O(R^{-2}).
\enn
The part with odd power of $\nu$ in the uniform expansion brings
the single linear on $R$ divergent contribution coming from
$\dot{S}^+_{-1}$:
\bd
2 (kR) \nu.
\ed
Therefore in the limit $R\to \infty$ the uniform expansion gives
the following divergent contribution:
\bd
2 (kR) \nu -2 \ln (kR) + \ln (ka).
\ed
Because later we have to take the derivative with respect to $k$
we may rewrite this expression in the following way
\bd
2 (kR) \nu  - \ln (ka).
\ed

To take the limit of large box in $S^+(+R) + S^-(-R)$ let us
reduce the radial equation to standard form of scattering problem
by changing the form of radial function $\phi \to \psi /r(\rho)$.
In this case the equation reads
\be\label{SchrEq}
\left[-\fr {d^2}{d\rho^2} + \fr {l(l+1)}{\rho^2 + a^2} + \fr {a^2
(1-2\xi)}{(\rho^2 + a^2)^2}\right] \psi = \lambda^2 \psi.
\ee

This equation looks similar to the equation of scattering problem in
one dimension (do not forget that $\rho\in (-\infty, +\infty)$)
with non-singular symmetric potential
\be\label{BarrierDef}
V_l (\rho,a)= \fr {l(l+1)}{\rho^2 + a^2} + \fr {a^2
(1-2\xi)}{(\rho^2 + a^2)^2}.
\ee

From standard theory of one-dimensional scattering we know that
there are two independent solutions which have the following
properties (as opposite to traditional notations $\phi (x)$ we use
solutions $\phi (-x)$ to make coincidence with functions we use)
\bnn
\phi_1(\rho) &\approx& \left\{\begin{array}{ll}
  s_{11}(\lambda) e^{-i\lambda\rho}, & \rho \to +\infty \adb\\
  e^{-i\lambda\rho} + s_{12}(\lambda) e^{i\lambda\rho}, & \rho \to -\infty
\end{array}, \right.\adb\\
\phi_2(\rho) &\approx& \left\{\begin{array}{ll} s_{21}(\lambda)
e^{-i\lambda\rho}+ e^{i\lambda\rho}, & \rho \to +\infty\adb\\
  s_{22}(\lambda) e^{i\lambda\rho}, & \rho \to -\infty
\end{array}, \right.
\enn
where $s_{\alpha\alpha }(\lambda)$ constitute the $s$ matrix of
scattering problem. Due to symmetry of the potential the
components of the matrix obey the relation $s_{22} = s_{11}$.

Now we change energy to imaginary axis: $\lambda \to ik\nu$ and
obtain
\bd
\phi_1(+R)\phi_2(-R)\strut_{R\to \infty} = s_{11}^2(ik\nu)
e^{2k\nu R}.
\ed
Therefore
\bd
\left[S^+(+R) + S^-(-R)\right]\strut_{R\to \infty} = \ln
(\phi_1(+R)\phi_2(-R))\strut_{R\to \infty} = \ln [s_{11}^2(ik\nu)]
+ 2kR\nu,
\ed
and the divergent parts in \Ref{Omega} are cancelled. Thus we
arrive at the following expression for $\Omega$
\bn\label{OmegaFin}
\Omega &=& \omega +  32\pi \sum_{l=0}^\infty \nu^{2}
\int_{\fr  1\nu}^\infty d y (y^2 - \fr {1}{\nu^2})^{1/2} \nonumber\adb\\
&\times&\fr {\partial}{\partial y} \left\{\ln [y s_{11}^2(iy\nu
m)] - \sum_{n=0}^2 \nu^{1-2n} \int_{-\infty}^{+\infty}
[\dot{S}_{2n-1}^+ - \delta_{0,n}]dx \right\}
\en

Thus, we express the finite part of the zero point energy in terms
of the $s$ matrix of scattering problem, namely in term of the
transmission coefficient of the barrier in imaginary axis. Similar
relation was found by Bordag in Ref.\cite{Bor95}. The potential
$V_l(\rho,a)$ of scattering problem has the following properties
\bn
V_l(0,a) &=& \fr {l(l+1) + 1 - 2\xi}{a^2} - {\rm the\ height\ of\
barrier},\nonumber\adb\\
\int_{-\infty}^{+\infty}V_l(\rho,a) d\rho &=&\fr {\pi[2l(l+1) + 1
- 2\xi]}{2a} - {\rm the\ work\ against\ potential\ barrier}.
\label{BarrierProp}
\en

Therefore the zero point energy has the form \Ref{Erenfin}, where
the function $\Omega$ is given by expression \Ref{OmegaFin}. We
should like to note that according with
\cite{BezBezKhu01,KhuSus02} the factor before logarithm term in
\Ref{Erenfin} is $(B_2 -B_1)_{R\to \infty}$. The origin of this
structure has been already noted in Eq. \Ref{LogTerms}.

Now we analyze qualitatively without exact numerical calculations
the behavior of energy for small and large radii of throat.
According with Eqs. \Ref{Small} and \Ref{Large} the zero point
energy in $3+1$ dimensions has the following behavior for small
and large values of throat
\bnn
E^{ren} &\approx& -\fr{B_2 \ln(am)^2}{32\pi^2},\ a\to 0, \adb\\
E^{ren} &\approx& -\fr{B_3}{32\pi^2 m^2},\ a\to \infty,
\enn
or in manifest form:
\bn
E^{ren} &\approx& -\fr  m{32} \fr {\ln (ma)^2}{ma} \left[3\xi^2 -
\xi + \fr  3{20}\right],\ a\to 0, \label{ato0}\\
E^{ren} &\approx& -\fr  m{32} \fr {1}{(ma)^3} \fr {1}{4032}
\left[5880\xi^3 - 6300\xi^2 + 2226\xi -257\right],\ a\to \infty.
\label{atoinfty}
\en
It is easy to verify that coefficient after logarithm in Eq.
\Ref{ato0}, which is contribution from $B_2$ in the limit $R\to
\infty$, is never to be zero or negative. It is always positive.
For this reason the zero point energy is positive for small radius
of throat for arbitrary constant of non-conformal coupling $\xi$.
In the domain of large radius of throat the expression in brackets
in Eq. \Ref{atoinfty}, which is the contribution from $B_3$ in the
limit $R\to \infty$, may change its sign. It is positive for $\xi
> 0.266$ (energy negative) and negative (energy positive) in the
opposite case. Therefore we conclude that there is minimum of
ground state energy if constant $\xi > 0.266$. The situation is
opposite to that which appeared in our last paper \cite{KhuSus02},
where the
energy for large value of throat (which was defined by $B_{5/2}$)
was always positive, but for small radius of throat it could change
its sign.

Let us now consider the semiclassical Einstein equations:
\be\label{ee}
G_{\mu\nu}=\fr{8\pi G}{c^4}\langle T_{\mu\nu}\rangle^{\rm ren},
\ee
where $G_{\mu\nu}$ is the Einstein tensor, and $\langle
T_{\mu\nu}\rangle$ is the renormalized vacuum expectation values
of the stress-energy tensor of the scalar field. The total energy
in a static space-time is given by
\bd
E=\int_V \varepsilon\, \sqrt{g^{(3)}} d^3x,
\ed
where $\varepsilon=-\langle T_{t}^{t}\rangle^{\rm ren}= - G_t^t
c^4/8\pi G$ is energy density, and the integral is calculated over
the whole space. In the spherically symmetric metric \Ref{Metric}
we obtain
\be\label{total}
E=-\fr{c^4}{2G} \int_{-\infty}^{\infty} G_t^t\, r^2(\rho) d\rho =
-\fr{c^4\pi a}{2G}.
\ee
The zero point energy has the following form
\be\label{ground}
E^{ren}=-\fr{\hbar c}{a}f(ma,\xi).
\ee
In the self-consistent case the total energy must coincide with
the ground state energy of the scalar field. Equating Eqs.
\Ref{ground} and \Ref{total} gives
\bd
\fr{c^4 a}{2G}=\fr{\hbar c}{a}f(ma,\xi),
\ed
or
\bd
a=l_P \sqrt{2f(ma,\xi)}.
\ed
Considering this equation at the minimum of zero point energy we
obtain some value of wormhole's radius. The concrete value of
radius may be found from exact numerical calculation of the zero
point energy as function of $ma$. But without this calculation we
conclude that the wormholes with the throat's profile \Ref{First}
may exist for $\xi >0.266$.

\section{The model of throat: $r(\rho) = \rho \coth\fr\rho\tau - \tau +
a$}\label{Model2}

We will not reproduce here the density for heat kernel
coefficients in manifest form due to their complexity. They may be
found from general formulas \Ref{hkcdensitygene}. There are two parameters
in this model $a$ and $\alpha  = \tau/a$. The
dimensional parameter $a$ characterizes this kind of wormhole as a whole.
Small value of this parameter indicates small size of wormhole. The
dimensionless parameter $\alpha $ characterizes the form of wormhole --
its ratio of the length and radius of throat. By changing the integration
variable $\rho = xa$ we observe that coefficients $B_2$ and $B_3$ have the
following structure which is clear from
dimensional consideration:
\bnn
B_2 &=& \int_{-\infty}^{+\infty} d\rho \overline{B}_2 = \fr 1a
\left[b_{2,2}\xi^2 + b_{2,1}\xi + b_{2,0}\right],\\
B_3 &=& \int_{-\infty}^{+\infty} d\rho \overline{B}_3 = \fr 1{a^3}
\left[b_{3,3}\xi^3 + b_{3,2}\xi^2 + b_{3,1}\xi + b_{3,0}\right],
\enn
where $b_{k,l}$ depend on the $\alpha  = \tau/a$, only. We note
that $b_{2,2}>0$ as it is seen from \Ref{OverB2}. Therefore we may
analyze the zero point energy for different values of the
parameter $\alpha $. From the general point of view we have the following
behavior of the zero point energy for small size of wormhole that is
for small value of $a\to 0$:
\be
E^{ren} \approx -\fr{\ln(ma)^2}{32\pi^2} B_2 = - \fr{\ln(
ma)^2}{32\pi^2 a}\left[b_{2,2}\xi^2 + b_{2,1}\xi + b_{2,0}\right].
\label{ErenSmall}
\ee

\begin{figure}
\centerline{\epsfxsize=8truecm\epsfbox{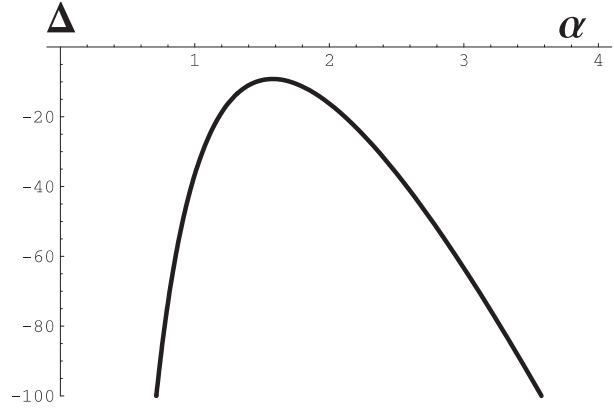}} \caption{The
discriminant $\Delta = b_{2,1}^2 - 4b_{2,2}b_{2,0}$ of polynomial
in $\xi$ as function of $\alpha $. It is always negative for all
values of $\alpha $. It means that the zero point energy is always
positive for small value of radius of throat.} \label{delta}
\end{figure}

Using the general expression for coefficient $B_2$ it is possible
to find in manifest form the polynomial in $\xi$ in
\Ref{ErenSmall} for great value of $\alpha \to \infty$ (small
radius of wormhole throat comparing with its length):
\be\label{Discsmall}
E^{ren} \approx -\fr{\sqrt{3\alpha } \ln(ma)^2}{240 a
}\left[30\xi^2 - 10\xi + 1\right],
\ee
and for small value of $\alpha \to 0$ (small length of wormhole
throat comparing with its radius):
\be\label{DiscLarge}
E^{ren} \approx -\fr{(15 - \pi^2)\ln(ma)^2}{1350\pi a \alpha  }
\left[240\xi^2 - 80\xi + 7\right].
\ee
The numerical calculations of the discriminant $\Delta = b_{2,1}^2 -
4b_{2,2}b_{2,0}$ of polynomial in $\xi$ as function of $\alpha $
is shown in Fig. \ref{delta}. From this figure and Eqs.
\Ref{Discsmall}, \Ref{DiscLarge} we conclude that the discriminant
is always negative for arbitrary values of $\alpha $. It means that
the zero point energy is always positive for small size of
wormhole for arbitrary constant of non-minimal coupling $\xi$ and
arbitrary ratio of the length of throat and its radius.

\begin{figure}
\centerline{\epsfxsize=10truecm\epsfbox{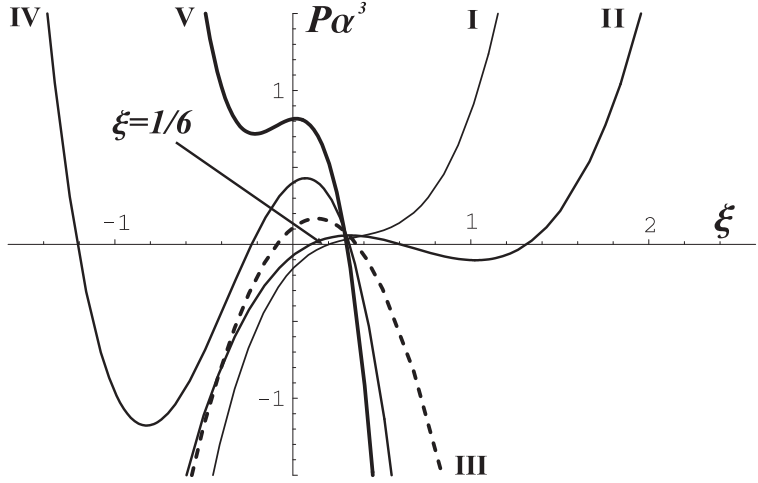}} \caption{The
plots of the polynomial $P\alpha ^3$ for different values of ratio
$\alpha  = \tau/a$. } \label{Steps}
\end{figure}

The behavior of zero point energy for large size of wormhole
($a\to\infty$) has the following form
\be
E^{ren} \approx -\fr{B_3}{32\pi^2 m^2} = -\fr{1}{32\pi^2 m^2 a^3}
\left[b_{3,3}\xi^3 + b_{3,2}\xi^2 + b_{3,1}\xi + b_{3,0}\right].
\ee
The zero point energy will get minimum for some value of $a$ if
above expression will be negative. Let us consider the polynomial
\be
P = b_{3,3}\xi^3 + b_{3,2}\xi^2 + b_{3,1}\xi + b_{3,0}
\ee
for different values of $\alpha $ starting from small value of it.
Zero point energy will get minimum if this polynomials is
positive. In the limit $\alpha \to 0$ we have approximately
\bnn
\alpha ^3 P&\approx& \frac{512\pi \left(45 - 4{\pi }^2
\right)}{135}\alpha {\xi }^3 - \left( \frac{256\pi \left(21 -
2{\pi }^2 \right)}{189} + \frac{256\pi \left(45 - 4{\pi
}^2 \right) }{135}\alpha  \right) {\xi }^2 \\
&+& \left( \frac{512\pi \left(21 - 2{\pi }^2 \right) }{945} +
\frac{224\pi \left(45 - 4{\pi }^2 \right) }{675}\alpha  \right)
\xi - \frac{368\pi \left( 21 - 2{\pi }^2 \right) }{6615} +
  \frac{44\pi \left( 45 - 4{\pi }^2 \right) }{2025}\alpha .
\enn
In this expression we saved terms up to $\alpha ^{-3}$. This
polynomial in the limit $\alpha \to 0$ has two complex roots and
one is real:
\bd
\xi\approx \frac{5\left( 21 - 2{\pi }^2 \right)}{14\left( 45 -
4{\pi }^2 \right)} \fr 1\alpha .
\ed
Because the coefficient with $\xi^3$ is positive the polynomials
will be positive for all
\bd
\xi > \frac{5\left( 21 - 2{\pi }^2 \right)}{14\left( 45 - 4{\pi
}^2 \right)} \fr 1\alpha .
\ed
Therefore for small values of $\alpha  = \tau/a$ we have
\textit{low} boundary for parameter $\xi$ where the wormhole may
exist (see Fig.\ref{Steps}(I)). The greater $\alpha $ the smaller
the low boundary of $\xi$. For $\alpha  > 1.136$ the conformal
connection $\xi = 1/6$ will be greater then the low boundary. At
the point $\alpha  = 1.26$ two domains appear where polynomial is
positive. First domain is $0.188 < \xi < 0.841$ and second $\xi >
0.841$ (see Fig.\ref{Steps}(II) for $\alpha  > 1.26$). The low
boundary of second domain will increase for greater $\alpha $ and
it disappears for $\alpha  = 1.65$. At this point the coefficients
$b_{3,3} = 0$ and the polynomial turns out to be parabola (see
Fig.\ref{Steps}(III)) with positive part in domain: $-0.088 < \xi
< 0.358$. For greater $\alpha $ we obtain \textit{upper} boundary
of $\xi$ where the polynomials is positive because the coefficient
with $\xi^3$ is negative. Starting from $\alpha  = 1.65$ we have
two domains where the polynomial is positive (see
Fig.\ref{Steps}(IV)). First one closes to $-0.088 < \xi < 0.358$
and another one is smaller then some negative value of $\xi$. For
$\alpha  = 2.08$ the high boundary of second domain will coincide
with low boundary of first domain and we get the only domain where
polynomial is positive $\xi < 0.309$. For greater value of $\alpha
$ this high boundary of $\xi$ tends to constant (see
Fig.\ref{Steps}(V)). Indeed, in the limit $\alpha \to \infty$ we
have
\be
P \approx \pi^2\sqrt{\fr{\alpha }{3}}\left[ -5\xi^3 + \fr 52\xi^2
- \fr 12\xi + \fr 1{21}\right]
\ee
and it is positive for all $\xi < 0.254$. We would like to note
that for $\alpha  > 1.136$ the polynomial is positive for $\xi =
1/6$.

Let us consider now what condition gives the Einstein's
equations. The energy corresponding for this configuration is
\be\label{total1}
E=-\fr{c^4}{2G} \int_{-\infty}^{\infty} G_t^t\, r^2(\rho) d\rho =
-\fr{2c^4}{G} \left[1 - \fr{15-\pi^2}{18} \alpha  \right]a.
\ee
Equating this energy with zero point energy
\bd
E^{\rm ren}=-\fr{\hbar c}{a}f(am,\alpha ,\xi)
\ed
we obtain the relation
\bd
\left[1 - \fr{15-\pi^2}{18}\alpha \right]a^2 = l_P^2f(am,\alpha
,\xi).
\ed
To find parameters of stable wormholes of this kind we have to
consider this equation at the minimum of function $f(am,\alpha
,\xi)$. Because the function $f(am,\alpha ,\xi)$ at the minimum is
positive, we conclude that the stable wormhole may exist for
$\alpha  < 18/(15 -\pi^2) =3.5$. For $\alpha  =3.5$ the polynomial
is equal to zero for $\xi = 0.278$. Therefore the stable wormholes
with this profile of throat may exist only for $\tau/a < 3.5$.
This upper boundary depends on the model of throat. For example,
wormhole with profile of throat
\bd
r(\rho) = \rho\tan \fr\rho\tau + a
\ed
gives another boundary, namely $\tau/a < 36/(\pi^2 - 6) = 9.3$.
Specific value of $\tau, a$ and region of $\xi$ may be found by
numerical calculation of the function $f(am,\alpha ,\xi)$.

\section{Conclusion}\label{Conc}

In the paper we analyzed the possibility of existence of the
semi-classical wormholes with metric \Ref{Metric} and throat's
profile given by Eqs.\Ref{First},\Ref{Second}. Our approach
consists of considering two heat kernel coefficients $B_2$ and
$B_3$. We developed a method for calculation the heat kernel
coefficient and obtained general expression for arbitrary
coefficient in background \Ref{Metric}. The first seven
coefficients in manifest form for arbitrary profile of throat are
given by Eqs. \Ref{hkcdensitygene} and \Ref{hkchalfgen}. The
sufficient condition for existence of wormhole is positivity of both
$B_2$ and $B_3$. Some additional conditions may follow from $t-t$
component of the Einstein's equations.

The common property of both models is that the zero point energy
for small size of wormhole is always positive for arbitrary
constant $\xi$. This statement is opposite to that obtained for
zero length throat model in Ref.\cite{KhuSus02}. The behavior of
zero point energy for large size of wormhole crucially depends on
the non-minimal coupling $\xi$ and parameters of the model. We
show that the wormholes with the first profile of throat may be a
self-consistent solution of semi-classical Einstein's equations if
the constant of non-minimal connection $\xi > 0.266$. This type of
wormholes is characterized by the only parameter $a$, which is the
radius of wormhole's throat. The space outside of throat
polynomially tends to Minkowkian and there is no way to define the length
of throat. We would like to note that the minimal connection
$\xi=0$ and conformal connection $\xi=1/6$ do not obey this
condition.

The second model of wormhole's throat \Ref{Second} is
characterized by two parameters $\tau$ and $a$. The latter is the
radius of wormhole's throat and first is the length of throat. It
is possible to introduce the length of throat because the space
outside the throat becomes Minkowskian exponentially fast. Suitable
illustration for this statement is Fig.\ref{Steps}(III). The
existence of this kind of wormholes crucially depends on the
parameter $\xi$ and ratio of the length and radius of throat: $\alpha
= \tau/a$. The general condition for $\alpha$ follows from
Einstein's equations, namely $\alpha < 3.5$. The wormhole with
very small parameter $\alpha$ may be self-consistent considered by
scalar massive field with large value of $\xi \sim 1/\alpha$. The
scalar field with conformal connection $\xi = 1/6$ may
self-consistently describe wormholes with $\tau/a \in (1.136,3.5)$.
For $\xi = 0$ we obtain another interval $\tau/a \in (1.473,3.5)$.

We would like to note that in the limit of zero length of throat
$\alpha =\tau/a \to 0$ there is no connection with results of our
last paper \cite{KhuSus02} where we considered wormhole with zero
length of throat. The point is that the model considered in that
paper is singular at the beginning. The scalar curvature is
singular at throat and there is a singular surface with
codimension one. For this case in Ref. \cite{GilKirVas01} the general
formulas for heat kernel coefficients were obtained which
can not be found as limit case of expression for smooth background
\cite{ZetaBook}. The reason of this lies in the following. The heat
kernel coefficients are defined as an expansion of heat kernel
over some dimension parameter which must be smaller then
characteristic scale of background. For smooth background we may
make this ratio small by taking appropriate value of expansion
parameter, but singular background has at the beginning the zero
value of background's scale. This leads to new form of heat kernel
coefficients. Furthermore, in the limit of large box $R\to\infty$ in
this background the coefficient $B_{5/2}$ survives and it
defines the behavior of energy for large size of wormhole.

Another interesting achievement of the paper is developing of the
zeta-function approach \cite{Bor95}. The radial equation in this
background \Ref{RadEqu} can not be solved in close form even for
simple profile of throat \Ref{First}. We obtain the general
formula for asymptotic expansion of solutions \Ref{ME}, using
which we find the heat kernel coefficients \Ref{hkcdensitygene},
\Ref{hkchalfgen} in general form. After renormalization the
zero point energy may be expressed in terms of the $S$ matrix of
scattering problem \Ref{Erenfin}, \Ref{OmegaFin}. More precisely,
we need only transmission coefficient $s_{11}$ of barrier
\Ref{BarrierDef}, \Ref{BarrierProp}. The point is that the radial
equation for massive scalar field in the background \Ref{Metric}
looks like a one-dimensional Schr\"odinger equation \Ref{SchrEq}
for particle with potential \Ref{BarrierDef}. This potential
depends on both orbital momentum $l$ of particle and non-minimal
coupling constant $\xi$ as well as on the radius of throat $a$ of
wormhole.

In first model the domain of $\xi$ for which the energy may
possess a minimum is limited from below. The reason of this
is connected with the fact that the effective mass
\bd
m^2_{eff} = m^2 + \xi \mathcal{R} = m^2 - \fr{2\xi a^2}{(\rho^2 +
a^2)^2}
\ed
may change its sign for some $\xi$ limited from below. The same
situation occurs in the short-throat flat-space wormhole
\cite{KhuSus02} where the scalar curvature is negative, too. This
is not the case for second model. The scalar curvature in this
model
\bd
\mathcal{R}=-\frac{6y^2 + 4\alpha  - 5{\alpha}^2 + \left( 4y^2 +
\alpha \left(-4 + 5\alpha \right)\right)\cosh (\frac{2y}{\alpha })
- 2y\left(-2 + 5\alpha \right) \sinh (\frac{2y}{\alpha })}{\tau^2
\sinh^4(\frac{y}{\alpha }){\left( 1 - \alpha  + y \coth
(\frac{y}{\alpha }) \right) }^2}
\ed
may change its sign depending on the parameters of model. For
small $\alpha$ it is negative but starting from $\alpha = 4/3$
the domain around $y=0$ appears, where the curvature becomes positive.
This domain becomes larger for larger values of $\alpha$ and for
$\alpha$ great enough the curvature is, in fact, positive. It is
in qualitative agreement with above consideration. Indeed for
small values of $\alpha$ (negative scalar curvature) we obtained
\textit{low} boundary for $\xi$ and vise versa, for large values of
$\alpha$ (positive scalar curvature) we obtained \textit{upper}
boundary for $\xi$.

\begin{acknowledgements}
The work was supported by part the Russian Foundation for Basic
Research grant N 02-02-17177.
\end{acknowledgements}

\appendix
\section{}\label{A}
In this Appendix we reproduce in manifest form some expressions,
which are rather long to reproduce them in the text. First of all
let us consider the first five terms of uniform expansion
\bnn
S^+(x) + S^-(-x) &=& \ln (\ak) - \fr  12 \ln (\dot{S}^2_{-1}
\rk^4) + \sum_{k=0}^\infty
\nu^{1-2k}\int_{-x}^{+x}\dot{S}^+_{2k-1}dx - \ln \left\{1 +
\sum_{k=1}^\infty
\nu^{-2k}\fr {\dot{S}^+_{2k-1}}{\dot{S}^+_{-1}}\right\}\adb\\
&=& \sum_{k=0}^\infty \nu^{1-2k}\int_{-x}^{+x}\dot{S}^+_{2k-1}dx +
\sum_{k=0}^\infty \nu^{-2k} E_{2k}. \nonumber
\enn
The coefficients with odd powers of $\nu$ read (we give the
integrands, only)
\bs
\bn\label{SeriesS}
\left[\nu^{1}\right] : \ \dot{S}_{-1}^+ &=& \sqrt{1 + \fr 1{\rk^2}}, \nonumber \adb\\
\left[\nu^{-1}\right] : \ \dot{S}_{1}^+  &=& -\fr{\xi \left( -1 +
\rk'^2 + 2\rk\rk''\right)}{\rk\sqrt{1 + \rk^2}} + \frac{ \left( -1
+ \rk'^2\right) + 2\rk\rk'' + 2 \rk^2 \left( -1 + 3\rk'^2 \right)
+ 6\rk^3\rk'' - \rk^4 + 4\rk^5\rk''}{8\rk \left( 1 + \rk^2 \right)^{5/2}},\nonumber\adb\\
\left[\nu^{-3}\right] : \ \dot{S}_{3}^+  &=& - \fr{\xi^2 {\left(
-1 + \rk'^2 + 2\rk\rk''\right) }^2}{2\rk\left( 1 + \rk^2
\right)^{3/2}} - \fr\xi{8\rk\left( 1 + \rk^2 \right)^{7/2}}\left[
-\left( -1 + \rk'^2 \right)^2 + 4\rk\left( 1 + \rk'^2 \right) \rk'' \right.\adb\nonumber\\
&-&\left.  2\rk^2\left( 1 - 8 \rk'^2 + 7\rk'(x)^4 - 2\rk''^2 -
8\rk'\rk^{(3)} \right) + 2 \rk^3\left( \left( 7 - 31\rk'^2
\right) \rk'' + 2\rk^{(4)} \right)\right.\adb\nonumber\\
&+& \left. \rk^4\left( -1 - 11 \rk'^2 + 12\rk'^4 - 4\rk''^2 +
12\rk'\rk^{(3)}\right) + 2 \rk^5\left( \left( 5 - 8\rk'^2
\right) \rk'' + 4 \rk^{(4)} \right)\right.\adb\nonumber\\
&-& \left. 4\rk^6\left( 2\rk''^2 + \rk'\rk^{(3)} \right) +
4\rk^7\rk^{(4)} \right] + \fr{1}{128\rk\left( 1 + \rk^2
\right)^{11/2}}\left[ - \left( -1 + \rk'^2\right)^2 + 4\rk\left(
1 + 3\rk'^2 \right)\rk'' \right.\adb\nonumber\\
&+& \left. 4\rk^2\left( -1 + 8\rk'^2 + 9\rk'^4 + 3\rk''^2 +
8\rk'\rk^{(3)}\right) + 4\rk^3\left( \left( 7 + 43\rk'^2 \right)
\rk'' + 2\rk^{(4)} \right) \right.\adb\nonumber\\
&+&\left. 2\rk^4\left( -3 + 17\rk'^2 - 294\rk'^4 + 40\rk''^2 +
88\rk'\rk^{(3)}\right) + \rk^5\left( 4\left( 15 - 136\rk'^2
\right)\rk'' + 40\rk^{(4)} \right) \right.\adb\nonumber\\
&+&\left. 4\rk^6\left ( -1 - 5\rk'^2 + 120\rk'^4 + 23\rk''^2 +
56\rk'\rk^{(3)}\right) + 4\rk^7\left( \left( 13 - 168\rk'^2
\right) \rk'' + 18\rk^{(4)} \right) \right.\adb\\
&-&\left. \rk^8\left( 1 + 24\rk'^2 + 8\rk''^2 -
48\rk'\rk^{(3)}\right) + 8\rk^9\left( 2\left( 1 + 2\rk'^2 \right)
\rk'' + 7\rk^{(4)} \right) - 32\rk^{10}\left( \rk''^2 +
\rk'\rk^{(3)}\right)  + 16\rk^{11}\rk^{(4)} \right],\nonumber
\en
and below are the coefficients with even power of $\nu$:
\bn\label{SeriesE}
\left[\nu^{0}\right] : \ E_0 &=& \ln (\ak) - \fr  12 \ln
(\dot{S}^2_{-1} \rk^4) \adb\nonumber \\
\left[\nu^{-2}\right] : \ E_2 &=& -\fr
{\dot{S}^+_1}{\dot{S}^+_{-1}}.
\en
\es

The functions $s_p$ are defined by relation
\bnn
\fr {\partial}{\partial k}(S^+(x) + S^-(-x)) &=&  \fr
{\partial}{\partial k}  \left\{ \sum_{p=0}^\infty
\nu^{1-2p}\int_{-kR}^{+kR}\dot{S}^+_{2p-1}(x)dx +
\sum_{k=0}^\infty \nu^{-2k} E_{2k} \right\}\adb\\
&=& \sum_{p=0}^\infty \nu^{1-2p} \int_{-R}^{+R}s_{2p-1}
(k\rho)d\rho + \sum_{p=0}^\infty \nu^{-2p}s_{2p}.
\enn
Below is the list of first four functions $s_{p}$ with odd
indices, (here $r = r(\rho)$ and $z= 1+ k^2 r^2(\rho)$)
\bs
\bn\label{Sodd}
s_{-1}&=& rz^{-1/2},\nonumber\adb\\
s_1&=& z^{-3/2}r\left[\xi \left( -1 + r'^2 + 2rr''\right) + \fr 18
\left\{1 - 4rr''\right\}  \right] + z^{-5/2}\fr{3r}{4}\left[-3r'^2
+ rr'' \right] + z^{-7/2}\left[\fr {25}8 rr'^2 \right],\nonumber\adb\\
s_3&=& z^{-5/2}\fr{3r}{2}\left[\xi^2 \left( -1 + r'^2 +
2rr''\right)^2 + \xi\fr 14\left\{ \left( -1 + r'^2\right) \left( 1
+ 12r'^2 \right) - 2r\left( -5 + 8 r'^2
\right)r''\right.\right.\nonumber\adb\\
&-& \left.\left. 4r^2\left( 2r''^2 + r'r^{(3)} \right) +
4r^3r^{(4)}\right\} + \fr 1{64} \left\{1 + 24r'^2 - 16r\left( 1 +
2r'^2 \right) r'' + 32r^2\left( r''^2 + r'r^{(3)}\right)  - 16r^3r^{(4)}\right\} \right]\nonumber \adb\\
&+& z^{-7/2} \fr{5r}{4}\left[\xi\left\{-19\left( -1 +
r'^2\right)r'^2  - 3r\left( 1 + 5r'^2 \right) r'' + 2r^2\left(
3r''^2 + 5r'r^{(3)} \right) \right\} \right.\nonumber\adb\\
&+&\left. \fr{1}{8}\left\{ - r'^2\left( 19 + 120r'^2 \right) +
r\left( 3 + 200r'^2 \right)r'' - 2r^2\left( 15r''^2 + 22r'r^{(3)}
\right)  + 2r^3r^{(4)}\right\} \right]\nonumber \adb\\
&+& z^{-9/2}\fr{175r}{8} \left[\xi r'^2\left( -1 + r'^2 +
2rr''\right) + \fr{1}{200} \left\{1014r'^4 + 38r^2r''^2 +
r'^2\left(25 - 832rr'' \right) + 56r^2r'r^{(3)}\right\} \right]\nonumber \adb\\
&+&z^{-11/2}\left[\fr{1989rr'^2}{32}\left( -3r'^2 +
rr''\right)\right] + z^{-13/2}\left[\fr{12155rr'^4}{128} \right],\nonumber\adb\\
s_5&=&z^{-7/2}\fr{5r}{2}\left[ \xi^3\left( -1 + r'^2 + 2rr''
\right)^3 + \fr{\xi^2}{8}\left\{ \left( -1 + r'^2\right)^2 \left(
3 + 112 r'^2 \right) + 12r\left( -1 + r' \right)\left( 1 +
r' \right) \left( 4 + 5r'^2 \right)r'' \right.\right.\adb\nonumber\\
&-&\left.\left. 4r^2\left( -27r''^2 + 48r'^2r''^2 - 26r'r^{(3)} +
26r'^3r^{(3)} \right) + 24r^3 \left( -2r''^3 - 2r'r''r^{(3)} -
r^{(4)} + r'^2r^{(4)} \right)\right.\right.\adb\nonumber\\
&+&\left.\left. 8r^4\left( 5r^{(3)}{}^2 + 6r''r^{(4)}
\right)\right\}  + \fr{\xi}{64}\left\{ \left( -1 + r'^2\right)
\left( 3 + 224r'^2 + 960r'^4 \right) - 2r\left( -39 - 860r'^2 +
1424r'^4 \right) r'' \right.\right.\adb\nonumber\\
&+&\left.\left. 8r^2\left( -54r''^2 + 186r'^2r''^2 - 69r'r^{(3)} +
80r'^3r^{(3)}\right) + 8r^3\left( 12r''^3 + 32r'r''r^{(3)} +
11r^{(4)}\right)\right.\right.\adb\nonumber\\
&-&\left.\left. 16r^4\left( 11r^{(3)}{}^2 + 15r''r^{(4)} +
3r'r^{(5)} \right) + 16r^5r^{(6)}\right\} + \fr{1}{512}\left\{1 +
112r'^2 + 960r'^4 - 4r\left( 9 + 424r'^2 + 192r'^4 \right)r''
\right.\right.\nonumber\adb\\
&+&\left.\left. 16r^2\left( 21r''^2 + 106r'^2r''^2 + 28r'r^{(3)} +
48r'^3r^{(3)} \right) - 64r^3 \left( 7r''^3 + 27r'r''r^{(3)} +
r^{(4)} + 6r'^2r^{(4)} \right)\right. \right.\nonumber\adb\\
&+&\left.\left. 32r^4\left( 9r^{(3)}{}^2 + 13r''r^{(4)} +
4r'r^{(5)} \right) - 32r^5r^{(6)}\right\}\right] + z^{-
9/2}\fr{7r}{8}\left[5\xi^2\left( -1 + r'^2 + 2rr''\right)\right.\nonumber \adb\\
&\times&\left.\left(-29r'^4 + r'^2\left( 29 - 15rr''\right) +
3rr''\left( -1 + 2rr''\right) + 20r^2r'r^{(3)} \right) +
\fr{\xi}{4}\left\{ -\left( -1 + r'^2\right) r'^2 \left( 145 +
2628r'^2 \right)\right.\right.\nonumber\adb\\
&+&\left.\left. r\left( -15 - 2839r'^2 + 2592r'^4 \right) r'' +
2r^2\left( 128r''^2 + 643r'^2r''^2 + 191r'r^{(3)} + 284r'^3r^{(3)}
\right)\right.\right.\nonumber\adb\\
&-&\left.\left. 2r^3 \left( 150r''^3 + 638r'r''r^{(3)} + 5r^{(4)}
+ 165r'^2r^{(4)}\right) + 8r^4\left( 7r^{(3)}{}^2 + 12r''r^{(4)} +
7r'r^{(5)} \right)\right\}\right.\nonumber\adb\\
&+&\left. \fr{1}{64}\left\{- r'^2\left( 145 + 5256r'^2 + 20160r'^4
\right)  + 3r\left( 5 + 1776r'^2 + 17952r'^4
\right) r'' \right.\right.\nonumber\adb\\
&-&\left.\left. 4r^2\left( 113r''^2 + 8440r'^2r''^2 + 166r'r^{(3)}
+ 3864r'^3r^{(3)}\right)  + 4r^3 \left( 676r''^3 +
2816r'r''r^{(3)} + 5r^{(4)} + 756r'^2r^{(4)}\right)\right.\right.\nonumber\adb\\
&-&\left.\left. 8r^4\left(58r^{(3)}{}^2 + 93r''r^{(4)} +
46r'r^{(5)} \right) + 8r^5r^{(6)}\right\} \right] +
z^{-11/2}\fr{9r}{16}\left[175r'^2\xi^2\left( -1 + r'^2 +
2rr''\right)^2\right.\nonumber\adb\\
&+&\left. \fr{\xi}{4}\left\{ \left( -1 + r'^2\right)r'^2 \left(
175 + 15054r'^2 \right) + 2rr'^2\left( 4413 + 1976r'^2 \right) r''\right.\right.\nonumber\adb\\
&-&\left.\left. 14r^2\left( 19r''^2 + 813r'^2r''^2 + 28r'r^{(3)} +
414r'^3r^{(3)} \right)  + 4r^3 \left( 133r''^3 +
638r'r''r^{(3)} + 221r'^2r^{(4)}\right)\right\}\right.\nonumber \adb\\
&+&\left.\fr{1}{64}\left\{r'^2\left( 175 + 30108r'^2 + 393408r'^4
\right)  - 8rr'^2\left( 2119 + 88298r'^2 \right) r''
\right.\right.\nonumber \adb\\
&+&\left.\left. 4r^2\left( 133r''^2 + 65998r'^2r''^2 +
196r'r^{(3)} + 31264r'^3r^{(3)} \right)  - 16r^3 \left( 628r''^3 +
2729r'r''r^{(3)} + 791r'^2r^{(4)} \right)\right.\right.\nonumber \adb\\
&+&\left.\left. 8r^4\left( 69r^{(3)}{}^2 + 110r''r^{(4)} +
54r'r^{(5)} \right)\right\} \right] + z^{- 13/2}\fr{11r}{32}
\left[ 221\xi r'^2\left\{ -37\left( -1 + r'^2\right)r'^2 -
9r\left( 1 + 5r'^2 \right)r''\right.\right.\nonumber\adb\\
&+&\left.\left. 2r^2\left( 9r''^2 + 5r'r^{(3)}\right)\right\} +
\fr{1}{8}\left\{- r'^4\left( 8177 + 332178r'^2 \right) +
rr'^2\left( 1989 + 396718r'^2 \right)r''\right.\right.\nonumber\adb\\
&-&\left.\left. 4r^2r'^2\left( 21015r''^2 + 10168r'r^{(3)} \right)
+ 2r^3 \left( 631r''^3 + 2782r'r''r^{(3)} + 815r'^2r^{(4)} \right)\right\} \right]\nonumber \adb\\
&+&z^{- 15/2}\fr{13rr'^2}{128}\left[ 12155\xi r'^2 \left( -1 +
r'^2 + 2rr''\right) + \fr{1}{8}\left\{ r'^2\left( 12155 +
2052348r'^2 \right) - 1484372rr'^2r''\right.\right.\nonumber \adb\\
&+&\left.\left. 4r^2\left( 34503r''^2 + 16880 r'r^{(3)}
\right)\right\}\right] + z^{-17/2}\left[\fr{3727125
rr'^4}{512}\left( -3r'^2 + rr''\right)\right] +
z^{-19/2}\left[\fr{7040125rr'^6}{1024} \right],
\en
and here is the list of first three functions $s_{p}$ with even
indices ($r = r(R)$ and $z= 1+ k^2 r^2(R)$)
\bn\label{Seven}
s_0 &=& -r^2 z^{-1},\nonumber\adb\\
s_2 &=& -z^{-2}2r^2\left[ \fr{1}{8}\left( 1 - 4 rr''\right)
+\xi\left( -1 + r'^2 + 2rr''\right)  \right]  - z^{-3}r^2\left[
-3r'^2 + rr''\right] -z^{-4}\left[\frac{15r^2r'^2}{4}
\right],\nonumber\adb\\
s_4 &=& -z^{-3}r^2\left[ 4\xi^2\left( -1 + r'^2 + 2rr''\right)^2 +
\xi \left\{\left( -1 + r'^2\right) \left( 1 + 6r'^2 \right)  -
2r\left( -4 + 5r'^2 \right) r'' - 2r^2\left( 4r''^2 +
r'r^{(3)}\right)  + 2r^3r^{(4)}\right\} \right.\nonumber\adb\\
&+&\left. \fr{1}{16}\left\{1 + 12r'^2 - 4r\left( 3 + 4r'^2
\right) r'' + 8r^2\left( 3r''^2 + 2r'r^{(3)}\right) - 8r^3r^{(4)} \right\} \right]\nonumber\adb\\
&-& z^{-4}3r^2\left[\xi \left(-11\left( -1 + r'^2\right)r'^2  -
2r\left( 1 + 5r'^2 \right) r'' + r^2\left( 4r''^2 +
5r'r^{(3)} \right)\right)\right.\nonumber\adb\\
&+& \left. \fr{1}{8}\left\{ -\left( 1 + 4r'^2 \right) \left( -1 +
15r'^2 \right)  +  2r\left( 1 + 53r'^2 \right)r'' +
r^2\left( -17r''^2 - 22r'r^{(3)}\right) + r^3r^{(4)} \right\}\right]\nonumber\adb\\
&-& z^{-5} r^2\left[ 30\xi r'^2\left( -1 + r'^2 + 2rr''\right) +
\fr{1}{4}\left\{ 3r'^2\left( 5 + 172r'^2 \right) - 432rr'^2r'' +
4r^2\left( 5r''^2 + 7r'r^{(3)}\right)\right\}  \right]\nonumber\adb\\
&-& z^{-6}\left[\fr{565r^2r'^2}{8}\left( -3r'^2 + rr''
\right)\right] - z^{-7}\left[\fr{1695r^2r'^4}{16}\right].
\en
\es
Below are the functions $\omega_a,\ \omega_b$ and $\omega_c$ with
definitions of the corresponding integrals.
\bn\label{Omegaabc}
\omega_a &=&\pi^2\left[-\fr  23 a + 32 a^3\int_0^1 \fr {d\nu\nu
}{e^{2\pi\nu a} + 1} \left(\sqrt{1-\nu^2} + \nu^2 \ln \left[\fr
\nu{1 + \sqrt{1- \nu^2}}\right]\right)\right],\nonumber\adb\\
\omega_b &=& -\fr  53 \pi^4 a^2 + a\pi^2\left[-16 (\gamma + 2\ln
2)\xi + \fr  43 (3 + 2\gamma)\right] - 8\pi^2 a \ln (2a)
(1-\tanh (\pi a)) (1 - 2\xi)\nonumber\adb\\
&-& 32 \pi^2 a\xi V_1 + \fr  43 \pi^2 a [V_1 + 6 V_2 - 4\pi a V_3
+ 2\pi a V_4 -  5\pi^2 a^2 V_5],\nonumber\adb\\
\omega_c &=&  U_1 + U_2 + U_3 + U_4 + U_5 + U_6.
\en

Here we introduced five functions for $\omega_b$
\bnn
V_1 &=& \fr {\pi a}2 \int_0^1 \fr {\ln (2a\nu)}{\cosh^2 (\pi
a\nu)} d\nu - \int_0^1 \fr {\nu d\nu}{e^{2\pi a\nu}+1} \left[\ln
\left(\fr \nu{1 + \sqrt{1-\nu^2}} \right)
+ \fr  1{1 + \sqrt{1-\nu^2}}\right],\adb\\
V_2 &=& \int_0^1 \fr {\nu d\nu}{e^{2\pi a\nu} + 1}\ln\left( \fr
\nu{1 + \sqrt{1-
\nu^2}}\right),\adb\\
V_3 &=& \int_0^1 \fr {d\nu}{\cosh ^2(\pi a\nu )}\ln\left( \fr
\nu{1 + \sqrt{1-
\nu^2}}\right),\adb\\
V_4 &=& \int_0^1 \fr {d\nu \sqrt{1-\nu^2}}{\cosh ^2(\pi a\nu )},\adb\\
V_5 &=& \int_0^1 \fr {\sinh (\pi a\nu )d\nu}{\cosh ^2(\pi a\nu )}
\left[\nu \sqrt{1-\nu^2} - \arccos\nu\right],
\enn
and six for $\omega_c$
\bnn
U_1&=& -\fr {229\pi^4}{2520} - \pi a\int_0^1 f_1(a\nu) \arccos \nu d\nu,\adb\\
U_2&=& \pi a\int_1^\infty f_3 (a\nu)\fr {d\nu}\nu + \pi a \int_0^1
\fr { f_3(a\nu)\nu d \nu}{1+ \sqrt{1-\nu^2}},\adb\\
U_3&=& \pi a\int_1^\infty f_5 (a\nu)\fr {d\nu}\nu \left(\fr  12 +
\fr  1{3\nu^2}\right) + \fr \pi 3 a \int_0^1
 f_5(a\nu)\nu d \nu \left[\fr  2{1+ \sqrt{1-\nu^2}} + \fr  1{2(1+ \sqrt{1-
\nu^2})^2}\right],\adb\\
U_4&=& \fr {\pi^2}2 \int_0^1 \fr {f_2d\nu}{\cosh^2 (\pi a\nu)} \ln
\left(\fr \nu {1+ \sqrt{1-\nu^2}}\right),\adb\\
U_5&=& \fr {\pi^2}4 \int_1^\infty \fr {f_4 d\nu}{\nu^2\cosh^2 (\pi
a\nu)} + \fr {\pi^2}4\int_0^1 \fr {f_4d\nu}{\cosh^2 (\pi
a\nu)}\left[ \ln \left(\fr \nu {1+ \sqrt{1-\nu^2}}\right) + \fr \nu {1+ \sqrt{1-\nu^2}}\right],\adb\\
U_5&=& \fr {\pi^2}8 \int_1^\infty \fr {f_6(\nu^2 +
1)d\nu}{\nu^4\cosh^2 (\pi a\nu)} + \fr {3\pi^2}{16}\int_0^1 \fr
{f_6d\nu}{\cosh^2 (\pi a\nu)}\left[ \ln \left(\fr \nu {1+
\sqrt{1-\nu^2}}\right) + \fr \nu {1+ \sqrt{1-\nu^2}} + \fr  13 \fr
\nu {(1+ \sqrt{1-\nu^2})^2}\right],
\enn
where
\bnn
f_1 (\nu)&=& \fr {4\pi}{315} (14595 \Pi_2(\nu) - 43638\Pi_3(\nu) +
47736
\Pi_4(\nu) - 17680 \Pi_5(\nu)),\adb\\
f_3 (\nu)&=& \fr {32\pi\xi}{3} (29 \Pi_2(\nu) - 10\Pi_3(\nu)) -
\fr {8\pi}{315}
(32445 \Pi_2(\nu) - 64113\Pi_3(\nu) + 55692 \Pi_4(\nu) - 17680 \Pi_5(\nu)),\adb\\
f_5 (\nu)&=& -\fr {64\pi\xi}{3} (16 \Pi_2(\nu) - 5\Pi_3(\nu)) +
\fr {32\pi}{315}
(6720 \Pi_2(\nu) - 10773\Pi_3(\nu) + 7956 \Pi_4(\nu) - 2210 \Pi_5(\nu)),\adb\\
f_2 (\nu)&=& -\fr {229\pi}{630},\adb\\
f_4 (\nu)&=& \fr {376\pi}{3}\xi-\fr {5948\pi}{315},\adb\\
f_6 (\nu)&=& 32\pi\xi^2 - \fr {544\pi}{3}\xi + \fr {8824\pi}{315},
\enn
and
\bnn
\Pi_2(\nu)&=& \fr {\pi^2}{2}Sc(\nu),\adb\\
\Pi_3(\nu)&=& \pi^2(Sc(\nu)+ \fr  18 \nu Sc'(\nu)),\adb\\
\Pi_4(\nu)&=& \pi^2(\fr  32 Sc(\nu)+ \fr {17}{48}\nu Sc'(\nu) +
\fr 1{48}\nu^2
Sc''(\nu)),\adb\\
\Pi_5(\nu)&=& \pi^2(2 Sc(\nu)+ \fr {259}{384}\nu Sc'(\nu) + \fr
{29}{384}\nu^2
Sc''(\nu))+ \fr {1}{384}\nu^3 Sc'''(\nu)),\adb\\
Sc(\nu)&=& \fr {\sinh (\pi a\nu )}{\cosh^3 (\pi a\nu)}.
\enn

\end{document}